\documentclass[11pt]{article}
\usepackage{jheppub}
\usepackage{amsmath,amsopn,amssymb,amsfonts,amsthm,mathrsfs,bbm,caption,latexsym,stmaryrd,verbatim,color,graphicx}
\usepackage{hyperref}
\usepackage[dvipsnames]{xcolor}

\let\la\langle
\let\ra\rangle
\let\l\left
\let\r\right
\let\rw\rightarrow

\let\wh\widehat
\let\mb\mathbb
\let\mc\mathcal

\newcommand{\nn}{\nonumber\\} 
\def\F#1#2{\,_{#1}F_{#2}}

\title{Irregular Singularities in the $H_3^+$ WZW Model}
\author[1]{D. Gaiotto}
\author[1,2]{J. Lamy-Poirier}
\affiliation[1]{Perimeter Institute for Theoretical Physics, %31 Caroline St. N., 
Waterloo, Ontario, Canada N2L 2Y5}
\affiliation[2]{Department of Physics and Astronomy,
University of Waterloo, Ontario, Canada N2L 3G1}

\abstract{We propose a definition of irregular vertex operators in the $H_3^+$ WZW model. Our definition is compatible with the duality \cite{Ribault:2005wp} between 
the $H_3^+$ WZW model and Liouville theory, and we provide the explicit map between correlation functions of irregular vertex operators in the two conformal field theories. 
Our definition of irregular vertex operators is motivated by relations to partition functions of N=2 gauge theory and scattering amplitudes in N=4 gauge theory}

\begin{document}

\maketitle
\flushbottom
\tableofcontents
%%%%%%%%%%%%%%%%%%%%%%%%%%%%%%%%%%%%%%%%%%%%%%%%%%%%%%%%%%%%

\newpage
\section{Introduction and Conclusions}
\label{sec:intro}

There are several protected quantities which are exactly computable in ${\cal N}=2$ four-dimensional gauge theories,
and have a rich physical and mathematical content. Many four-dimensional gauge theories can be engineered from the twisted compactification of a six-dimensional CFT on a Riemann surface \cite{Witten:1997sc,Gaiotto:2009hg,Gaiotto:2009we}.  Protected quantities associated to such ``class {\cal S}'' gauge theories can be usually given an interpretation in terms of mathematical structures attached to the Riemann surface. In particular, the $\Omega$-deformed partition function of the four-dimensional theories takes the form of two-dimensional conformal blocks for Virasoro or W-algebras \cite{Alday:2009aq,Wyllard:2009hg}.
Other current algebras arise from modifications of the four-dimensional setup: super-Virasoro and para-Virasoro (or W-algebras) \cite{Nishioka:2011jk,Bonelli:2011jx} arise from geometric quotients of the $\Omega$ background, WZW current algebras and their Hamiltonian reductions arise in the presence of extra surface defects \cite{Alday:2010vg,Kozcaz:2010yp}.
Furthermore, the partition function on a four-dimensional ellipsoid \cite{Pestun:2007rz,Hama:2012bg} coincides with the correlation functions of the Liouville-like conformal field theories directly associated to the appropriate current algebra: Liouville and Toda, super or para-Liouville or super or para-Toda, the $H_3^+$ WZW model and its higher rank $SL(N,\mathbb{C})/SU(N)$ generalizations. 

There is an intricate dictionary which pairs up a specific class {\cal S} theory with a specific
choice of correlation function in the two-dimensional conformal field theories. 
Standard superconformal field theories in the class {\cal S} can be matched to generic 
correlation functions of standard vertex operators in the two-dimensional CFTs. 
Asymptotically free gauge theories and Argyres-Douglas-like theories
are matched with conformal blocks and correlation functions involving 
more exotic vertex operators, dubbed irregular vertex operators \cite{Gaiotto:2009hg,Gaiotto:2009ma,Gaiotto:2012sf}.   
Indeed, asymptotically free gauge theories and Argyres-Douglas-like theories
can be derived by a careful degeneration limit of 
the standard superconformal field theories. Correspondingly, on the two-dimensional CFT side,
irregular vertex operators arise from a collision of standard vertex operators, whose conformal 
dimension is sent to infinity in a specific way. 

The definition of irregular vectors has been given systematically 
for the Virasoro current algebra and Liouville theory correlation functions in \cite{Gaiotto:2012sf}.
The main purpose of this note is to give a systematic definition of irregular conformal blocks
for the $\hat sl(2)$ current algebra and $H_3^+$ WZW theory. Similar, but sometimes more restrictive 
definitions appeared before in the mathematical literature, see for example \cite{Nag1, Nag2, Nag3} or 
\cite{FFT}. 

We have a good reason for picking this example, among all possible generalizations of the Virasoro 
problem: it is a toy model for a different, deep problem 
which arises in the calculation of scattering amplitudes at strong coupling in planar ${\cal N}=4$ SYM theory. 
Scattering amplitudes in planar ${\cal N}=4$ SYM theory can be related to 
correlation functions of polygonal Wilson loops with null edges \cite{Alday:2009yn,Alday:2009ga,Alday:2009dv,Alday:2010vh}. At strong coupling, 
the Wilson loop correlation function should be computable in terms of a string world-sheet
in $AdS_5$, ending on the null polygon at the boundary. At leading order, the calculation is classical, and one has to compute the area of a minimal-area surface 
bounded by the null polygon. In order to go beyond the leading order, one has to do a full quantum computation 
on the world-sheet theory describing a superstring moving in $AdS_5 \times S^5$. In particular, 
one has to find out how to impose appropriate boundary conditions at infinity, 
encoding in the full quantum theory the shape of the null Wilson loop at the boundary.

In the classical problem, the boundary condition at infinity produces an ``irregular singularity'' 
in a certain auxiliary connection, which lead to Stokes phenomena. The Stokes data at the singularity 
encodes the shape of the null polygon. Thus one needs to define a vertex operator for the 
world-sheet theory, which inserted at infinity leads to similar singularities and Stokes phenomena. 
Irregular vectors in Liouville theory give rise to irregular singularities in 
the differential equations satisfied by degenerate fields, and to Stokes phenomena. 
It is our hope that irregular singularities in the $H_3^+$ WZW theory
may provide a closer analogue to the irregular singularities which are needed 
in the world-sheet theory of a superstring moving in $AdS_5 \times S^5$.

In this paper we derive the Ward identities which define irregular vertex operators and conformal blocks for the $\wh{sl}(2)$ current algebra. 
We define a collision limit for $r+1$ standard highest weight vertex operators, which leads to a generic rank $r$ irregular vertex operator. 
Our results generalize in a natural way the definitions given for the Virasoro algebra in \cite{Gaiotto:2012sf}, and allow us to formulate a simple conjecture on the 
structure of Ward identities for irregular vertex operators in a general current algebra which admits a free field realization. 
We also give a generalization of the known duality between $H_3^+$ WZW correlation functions and Liouville theory correlation functions, 
which includes irregular vertex operators on both sides of the relation. This duality can be used in conjunction with the results of \cite{Gaiotto:2012sf} 
in order to understand the structure of WZW conformal blocks with irregular singularities. 
Finally, we derive the KZ equations satisfied by irregular conformal blocks and correlation functions, and we give a semiclassical description 
of irregular vertex operators. 

While this paper was in preparation, we received the paper \cite{Kanno:2013vi}, which has substantial overlap with our section \ref{sec:Wthree}

\section{Irregular vectors and current algebras}
\label{sec:current}

In this section we give the explicit current algebra Ward identities for irregular vectors. Our main result in this section is a parameterization of the singularity in the currents which is compatible with the current algebra commutation relations, and with collision limits. In section \ref{sec:Virasoro} we review the Virasoro algebra analysis from \cite{Gaiotto:2012sf}. In section \ref{sec:sl2} we look at the collision limit for several colliding primary fields of a $\wh{sl}(2)$ current algebra. We look for an appropriate set of parameters which can be kept independent of the collision parameters while giving rise to a finite limit. The final form of the irregular Ward identities parameterizes the singularity of the currents by a set of bosonic oscillators, with a structure which mimics the Wakimoto free-field representation. The simplicity of our result suggests a natural extension to generic current algebras which admit a free field realization, which we discuss in section \ref{sec:Wthree}.

\subsection{The Virasoro case}
\label{sec:Virasoro}
The Virasoro Ward identity for irregular vertex operators has been studied in \cite{Gaiotto:2012sf}. An irregular vertex operator of rank $r$ is defined as an eigenvector of the Virasoro modes $\{L_{2r},\cdots,L_r\}$, annihilated by the higher modes $L_n$, $n>2r$:
\begin{align}\label{IrregDef}
L_m\Psi_{\bf\Lambda}^{(r)}=\l\{\begin{array}{ll}\Lambda_m\Psi_{\bf\Lambda}^{(r)}&2r\geq m\geq r\\
0&m>2r
\end{array}\r.,
\end{align}
where $\Lambda$ is the set of eigenvalues ${\bf\Lambda}=(\Lambda_{r},\cdots,\Lambda_{2r})$. The Virasoro algebra forbids the field to be an eigenvector of any other mode. The Virasoro Ward identity for a rank $r$ field $\Psi_{\bf\Lambda}^{(r)}$ can thus be written in the form
\begin{align}
T(w)\Psi_{\bf\Lambda}^{(r)}(z)\sim \l[\sum_{k=r}^{2r}\frac{\Lambda_k}{(w-z)^{k+2}}+\sum_{k=0}^{r-1}\frac{\mc L_k}{(w-z)^{k+2}}+\frac{\partial_z}{(w-z)}\r]\Psi_{\bf\Lambda}^{(r)}(z).
\end{align}
where the differential operators $\mc L_k$ need to be compatible with the Virasoro algebra. The general solution proposed in \cite{Gaiotto:2012sf} requires one to 
express the eigenvalues $\Lambda_k$ in terms of a new set of auxiliary parameters ${\bf c}=(c_0=\alpha,c_1,\cdots,c_r)$
\begin{align}
\Lambda_k=(k+1)Q c_k-\sum_{l=0}^{k}c_lc_{k-l},
\end{align}
where $c_{n>r}\equiv0$. This is a generalization of the parametrization $\Delta=\alpha(Q-\alpha)$ familiar from Liouville theory. In terms of these parameters, the differential operators read
\begin{align}
\mc L_k=(k+1) Q c_k-\sum_{l=0}^{k}c_lc_{k-l}+\sum_{l=k+1}^{r}(l-k)c_l\partial_{c_{l-k}}.
\end{align}
(Note that the derivatives $\partial_{c_k}$ are only defined for $1\leq k\leq r$.) This allows to write the Ward identity in the form
\begin{align}\label{VirasoroLiouville}
T(w)\Psi_{\bf\Lambda}^{(r)}(z)\sim \l[\sum_{k=0}^{2r}\frac{(k+1) Q c_k-\sum_{l=0}^{k}c_lc_{k-l}+\sum_{l=k+1}^{r}(l-k)c_l\partial_{c_{l-k}}}{(w-z)^{k+2}}+\frac{\partial_z}{(w-z)}\r]\Psi_{\bf\Lambda}^{(r)}(z).
\end{align}

The parameterization of the Virasoro Ward identities is clearly inspired to the free-field realization of the Virasoro algebra, but it is conceptually distinct. 
The correct statement is that an irregular vertex operator for the Virasoro algebra is an object which satisfies the same Ward identities as a coherent state 
\begin{equation}
\exp \left[ \sum_k c_k a_{-k} \right] |0 \rangle
\end{equation}
in the free field theory which realizes the Virasoro algebra. The $a_{-k}$ are the creation modes of the free scalar. 
In general, there is a linear space of solutions of the irregular Virasoro Ward identities, i.e. irregular conformal blocks. 
As for standard conformal blocks, the naive free-field description only provides a special solution in that space, and 
intricate configurations of screening charges are required to give a free-field description of general solutions.

A more intrinsic way to understand the origin of the irregular Virasoro Ward identities and the meaning of irregular conformal blocks 
is to define a rank $r$  irregular vertex operator as a collision limit of $r+1$ regular vertex operators. 
Starting from the Ward identity
\begin{equation}
T(w) \prod_j \Psi_{\alpha_j}(z_j) \sim \sum_i \left[  \frac{\alpha_i(Q-\alpha_i)}{(w-z_i)^2} + \frac{1}{w-z_i} \partial_{z_i} \right] \Psi_{\alpha_j}(z_j)
\end{equation}
and defining \footnote{Note that this rescaling allows to obtain a finite limit for the Virasoro Ward identities. 
Virasoro conformal blocks will have a finite limit if properly normalized and Liouville theory correlation functions 
will also have a finite limit (see \cite{Gaiotto:2012sf} for details). In order to extend the collision limits to correlation functions of other CFTs, one may need 
to add further pre-factors which take into account the behaviour of three-point functions and the normalization conventions for the vertex operators in that CFT.}
\begin{equation}
\Psi = \prod_{k,t} (z_k - z_t)^{2 \alpha_k \alpha_t} \prod_j \Psi_{\alpha_j}(z_j) 
\end{equation}
we can bring the $z_i$ to a common point $z$, while the $\alpha_i$ are sent to infinity in such a way that 
\begin{equation}
\sum_i \frac{\alpha_i}{w-z_i} \to \sum_k \frac{c_k}{(w-z)^{k+1}} 
\end{equation}
This collision limit brings the Ward identities for $\Psi$ to the ones for the irregular vector $\Psi_{\bf\Lambda}^{(r)}(z)$.

\subsection{Irregular $\wh{sl}(2)$ currents}
\label{sec:sl2}
The Ward identity for a standard spin $j$ primary field  the $\wh{sl}(2)$ valued current $J(w)=J^a(w)t_a$ reads 
\begin{align}\label{WardJ}
J^a(w)\Phi_j(\mu|z)\sim\frac{1}{w-z}\mc D^a\Phi_j(\mu|z).
\end{align}
where $\mc D^a$ are spin-$j$ generators for $sl(2)$. We use the following realization of the generators: \footnote{Here the fields are rescaled by a factor $\mu^{j}$ relative to the ones in \cite{Ribault:2005wp}, so the differential operators are slightly different.}
\begin{align}
\mc D^-=\mu,\qquad
\mc D^0=j-\mu\partial_\mu,\qquad
\mc D^+=2j\partial_\mu-\mu\partial_\mu^2.
\end{align}
Here the fields are related to the more common $x-$basis \cite{Alday:2010vg,CFT,Teschner:1997fv,Teschner:1997ft} by a Fourier transform and a rescaling \cite{Ribault:2005wp}
\begin{align}
\Phi_j(\mu|z)=\frac1\pi|\mu|^{4j+2}\int_{\mb C}d^2x~e^{\mu x-\bar\mu\bar x}\Phi_j(x|z)
\end{align}
In that basis the fields $\Phi_j(x|z)$ satisfy a Ward identity similar to (\ref{WardJ}), where the differential operators are replaced by $\tilde{\mc D}^a$:
\begin{align}
\tilde{\mc D}^-=-\partial_x,\qquad \tilde{\mc D}^0=x\partial_x-j,\qquad \tilde{\mc D}^+=x^2\partial_x-2jx.
\end{align}
The use of the $\mu$ basis is rather convenient both for the collision limit, and to make contact with the Liouville-$H_3^+$ duality.

It will be useful to remember how such Ward identities arise in the context of a Wakimoto free-field realization of the current algebra \cite{CFT} 
\begin{align}\label{Wakimoto}
J^-(w)=&\beta(w),\nonumber\\
J^0(w)=&\partial\phi(w)+(\gamma\beta)(w),\nonumber\\
J^+(w)=&-2(\partial\phi\gamma)(w)-k\partial\gamma-(\beta(\gamma\gamma))(w),
\end{align}
by looking at a primary field $E_j(\mu|z)$ for the scalar $\phi$ and the $\beta \gamma$ system. 
\begin{align}
\beta(w)E_j(\mu|z)&\sim\frac{\mu}{w-z}E_j(\mu|z),\nonumber\\
\partial\phi(w)E_j(\mu|z)&\sim\frac{j}{w-z}E_j(\mu|z),\nonumber\\
\gamma(w)E_j(\mu|z)&\sim-\partial_\mu E_j(\mu|z).
\end{align}
Our final answer for the irregular vector Ward identities will coincide with the Ward identities satisfied by an appropriate coherent state $E^{(r)}_{\bf j}({\boldsymbol \mu}|z)$ for $\phi$ and the $\beta \gamma$ system.
\begin{align}\label{WakimotoWard}
\beta(w)E^{(r)}_{\bf j}({\boldsymbol \mu}|z)\sim&\sum_{m=0}^{r}\frac{\mu_m}{(w-z)^{m+1}}E^{(r)}_{\bf j}({\boldsymbol \mu}|z),\nonumber\\
%\frac{i}{\sqrt{2}\alpha_+}
\partial\phi(w)E^{(r)}_{\bf j}({\boldsymbol \mu}|z)\sim&\sum_{m=0}^{r}\frac{j_m}{(w-z)^{m+1}}E^{(r)}_{\bf j}({\boldsymbol \mu}|z),\nonumber\\
\gamma(w)E^{(r)}_{\bf j}({\boldsymbol \mu}|z)\sim&-\sum_{m=0}^{r}(w-z)^{m}\partial_{\mu_m}E^{(r)}_{\bf j}({\boldsymbol \mu}|z),
\end{align}

Combining these OPE with the Wakimoto realization we arrive to our proposal for the irregular Ward identities for an irregular vector of rank $r$
$\Phi^{(r)}_{\bf j}({\boldsymbol \mu}|z)$, labeled by ${\bf j}=\{j_n\}$, ${\boldsymbol \mu}=\{\mu_n\}$, with  $0\leq n\leq r$.
\begin{align}\label{IrregCurrents}
J^-(w)\Phi^{(r)}_{\bf j}({\boldsymbol \mu}|z)\sim&\sum_{m=0}^{r}\frac{\mu_m}{(w-z)^{m+1}}\Phi^{(r)}_{\bf j}({\boldsymbol \mu}|z),\nonumber\\
J^0(w)\Phi^{(r)}_{\bf j}({\boldsymbol \mu}|z)\sim&\sum_{m=0}^{r}\frac{j_m-\sum_n\mu_n\partial_{\mu_{n-m}}}{(w-z)^{m+1}}\Phi^{(r)}_{\bf j}({\boldsymbol \mu}|z),\nonumber\\
J^+(w)\Phi^{(r)}_{\bf j}({\boldsymbol \mu}|z)\sim&\sum_{m=0}^{r}\frac{2\sum_nj_n\partial_{\mu_{n-m}}-\sum_{n,p}\mu_n\partial_{\mu_{p}}\partial_{\mu_{n-p-m+1}}}{(w-z)^m}\Phi^{(r)}_{\bf j}({\boldsymbol \mu}|z).
\end{align}
As in the Virasoro case, the coherent state only provides a model to the Ward identities, but not the general solution of the Ward identities. 

We obtain the same result if we define a rank $r$ WZW irregular vertex operator in terms of the collision of 
 $r+1$ regular vertex operators. For this purpose, consider $r+1$ primary fields $\Phi_{j^k}(\mu^k|z_k)$, using an upper index for $j^k$ and $\mu^k$ to distinguish from the mode index. As each field approach the collision point $z$, we allow $j^k$ and $\mu^k$ to diverge in the collision limit $q_k\rw 0$, but we require the current $J^a(w)$ to remain finite. 
If we adjust the $\mu^k$ in such a way that 
\begin{equation}
\sum_i\frac{\mu^i}{w-z_i} \to \sum_{m=0}^{r}\frac{\mu_m}{(w-z)^{m+1}}
\end{equation} 
then the Ward identity for $J^-(w)$  
\begin{align}
J^-(w)\prod_k\Phi^{(r)}_{j^k}({\mu^k}|z_k)\sim \sum_i\frac{\mu^i}{w-z_i}\prod_k\Phi^{(r)}_{j^k}({\mu^k}|z_k)
\end{align}
will obviously have the correct limit. 

Note that the change of variables from $\mu^k$ to $\mu_m$ is generated by the Vandermonde matrix
\begin{align}
\mu_{m-1}=\sum_k M_{mk}\mu^k,\quad\text{where}\quad M_{mk}=(z_k-z)^{m-1}
\end{align}
From the change of variables we read the relation between the derivatives, $\partial_{\mu^k} = \sum_{m=1}^{r+1} M_{mk}\partial_{\mu_{m-1}}$, and a straightforward calculation shows that the contribution from the $-\mu\partial_{\mu}$ terms in $J^0(w)$ has a finite limit by itself. If we adjust the $j^k$ in such a way that 
\begin{equation}
\sum_i\frac{j^i}{w-z_i} \to \sum_{m=0}^{r}\frac{j_m}{(w-z)^{m+1}}
\end{equation} 
then we arrive to the desired $J^0(w)$ Ward identity. Finally, some tedious algebra shows that the $J^+(w)$ Ward identities have a finite limit, and we recover the 
Ward identities for a rank $r$ WZW irregular vector. 
 Notice that the relation between derivatives tells us how one would do the collision limit in the $x$ basis: $x^k = \sum_{m=1}^{r+1} M_{mk}x_{m-1}$, i.e. the $x^k$ parameters collide in a similar pattern as the $z_k$. The irregular vertex operators in the $x_m$ and $\mu_m$ bases are again related by Fourier transform and a rescaling.

The WZW Ward identities are not sensitive to a rescaling by a function of the $z_k$ in the collision limit. Such a factor can be fixed by requiring the stress tensor Ward identities to also have a specific finite limit. Choices which differ by a finite function of the $j_m$ in the limit will give slightly different, albeit equivalent forms for the Virasoro Ward identities. A natural choice for the Ward identities is to mimic the form of the stress tensor in the Wakimoto free field realization of the theory.
\begin{equation}
T=-b^2(\partial\phi\partial\phi)+b^2\partial^2\phi-(\beta\partial\gamma)
\end{equation}
This expression actually allows to write the whole Virasoro Ward identity, since the first $r$ regular modes of $\partial\phi$ are fixed by its mode algebra. The result takes the form (\ref{VirasoroLiouville}) with $c_m \to bj_m$, $Q \to -b$, 
plus $\mu\partial\mu$ terms which arise from the $\beta\partial\gamma$ term: 
\begin{align}\label{VirasoroWard}
T(w)\Phi^{(r)}_{\bf j}({\boldsymbol \mu}|z)
\sim&
\l(-b^2\sum_{m=0}^{2r}\frac{\sum_{n=0}^{m}j_n j_{m-n}+(m+1)j_{m}
}{(w-z)^{m+2}}+\frac{\partial_z}{(w-z)}\r.\nn
&\qquad+\l.\sum_{m=0}^{r-1}\frac{\sum_{n=1}^{r}n(\mu_{n+m}\partial_{\mu_n}+j_{n+m}\partial_{j_n})}{(w-z)^{m+2}}\r)\Phi^{(r)}_{\bf j}({\boldsymbol \mu}|z).
\end{align}
In appendix \ref{app:sl2Vir}, we derive the same result from the collision limit (in a slightly modified form). It is an important building block for the generalization of the KZ equation, which is derived in appendix \ref{app:KZ}.

\subsection{Generalization to other current algebras}
\label{sec:Wthree}
It is pretty natural to take the Virasoro and $\wh{sl}(2)$ irregular Ward identities as an example of a general proposal on how to describe irregular vectors for any current algebra which admits a free-field realization: mimic the Ward identities satisfied by an appropriate coherent state for the free fields. For finite Lie algebras, the current algebra can be obtained from a Wakimoto realization, in a way similar $\wh{sl}(2)$. In this case we have one pair of bosonic ghosts $\beta^a$, $\gamma^a$ per positive root, and one free boson $\partial\phi^i$ per element of the Cartan subalgebra. The irregular Ward identities for these fields work exactly in the same way as in the $\wh{sl}(2)$ case, so the current Ward identities can be obtained by their Wakimoto representations analogously. For example, since the energy-momentum tensor in the Wakimoto representation is simply the sum of the one for the free fields, its irregular expression has to take the form
\begin{align}
T(w)\Phi^{(r)}_{\bf j}({\boldsymbol \mu}|z)
\sim&
\l(-b^2\sum_{m=0}^{2r}\frac{\sum_{n=0}^{m}{\boldsymbol j_n}\cdot{\boldsymbol j_{m-n}}+(m+1){\boldsymbol\rho}\cdot {\boldsymbol j_{m}}
}{(w-z)^{m+2}}+\frac{\partial_z}{(w-z)}\r.\nn
&\qquad\l.+\sum_{m=0}^{r-1}\frac{\sum_{n=1}^{r}n(\sum_a\mu^a_{n+m}\partial_{\mu^a_n}+\sum_i j^i_{n+m}\partial_{j^i_n})}{(w-z)^{m+2}}\r)\Phi^{(r)}_{\bf j}({\boldsymbol \mu}|z),
\end{align}
where ${\boldsymbol\rho}$ is the Weyl vector. (As in the $\wh{sl}(2)$ case, this expression is not unique, as shifting the field by a function of the parameters $j_m^i$ can modify the expression slightly.)

While the use of free-field realizations works well for affine Lie algebras, it can be used for other current algebras as well. We will give here a simple example: irregular vectors for the $\mc W_3$ algebra. This algebra has two generators: the energy-momentum tensor $T(z)$, and a spin 3 field $W(z)$. A free-field realization of this algebra is obtained from a triplet of free bosons $\phi_i$, constrained by $\sum_i\phi_i=0$. In this realization the currents take the form \cite{Kanno:2009ga}
\begin{align}
T&=-\frac12(\partial\phi_1^2+\partial\phi_2^2+\partial\phi_3^2)+Q(\partial^2\phi_3-\partial^2\phi_1)
,\nn
W&=
i\sqrt\frac3{30Q+8}\l(\partial\phi_1^3+\partial\phi_2^3+\partial\phi_3^3-Q^2\partial^3\phi_2-2QM_{ij}\partial\phi_i\partial^2\phi_j\r)
\end{align}
where $Q$ is a parameter of the theory. The matrix $M$ has the nonzero entries $M_{12}=M_{13}=M_{23}=M_{33}=-M_{11}=1$
Primary fields in this representation satisfy 
\begin{align}
\partial\phi_i (w)V_{\boldsymbol c}(z)\sim\frac{c^i}{w-z}V_{\boldsymbol c}(z),
\end{align}
for $i=1,2$. Following the same method as before, we can guess that irregular vectors will satisfy
\begin{align}
\partial\phi_i (w)V^{(r)}_{\boldsymbol c}(z)\sim\l(\sum_{m=0}^r\frac{c^i_m}{(w-z)^{m+1}}
+\sum_{m=1}^r m \partial_{c^i_m}(w-z)^{m-1}+\mc O((w-z)^r)\r)V^{(r)}_{\boldsymbol c}(z),
\end{align}
which implies the irregular currents

\begin{align}
T(w)&V^{(r)}_{\boldsymbol c}(z)\sim 
\l(\sum_{m=0}^{2r}\frac{-\frac12\sum_{i,n}{c^i_n}{c_{m-n}^i}+(m+1)Q(c_{m}^3-c_m^1)+\sum_{i,n}nc^i_{n+m}\partial_{c^i_n}
}{(w-z)^{m+2}}\r.\nn
&\l.+\frac{\partial_z}{(w-z)}\r)
V^{(r)}_{\boldsymbol c}(z),\nn
W(w)&V^{(r)}_{\boldsymbol c}(z)\sim i\sqrt\frac3{30Q+8}
\l(\sum_{m=r}^{3r}\frac{\sum_{i,n,p}c_n^ic_p^ic_{m-n-p}^i+\sum_{i,n,p}c_n^ic_p^i\partial_{c_{n+p-m}^i}
}{(w-z)^{m+3}}\r.\nn
&\l.-\frac{
2Q\sum_{i,j,n}(n+1)M_{ij}c_n^ic_{m-n}^j+Q^2(m+1)(m+2)c_m^i
}{(w-z)^{m+3}}
+\mc O((w-z)^{-r-2})\r)
V^{(r)}_{\boldsymbol c}(z).
\end{align}
The Ward identities do not fix several singular terms in the OPE of $W$. This is not a surprise: the Ward identities for a regular vector $V$ of the W-algebra contain descendants 
such as $W_{-1} V$ and $W_{-2} V$ which cannot be rewritten in terms of derivatives in the parameters of the conformal block. The Ward identities for a rank $r$ irregular vector contain 
$r+2$ such descendants. 

\section{Correlation functions}
Our discussion of the collision limits until this point only concerned the holomorphic (or anti-holomorphic) Ward identities. In the Virasoro case, there is strong evidence \cite{Gaiotto:2012sf} that the collision limit is also sensible at the level of full correlation functions for Liouville theory. The Liouville theory correlation functions are assembled from DOZZ three-point functions, \cite{Dorn:1992at,Dorn:1994xn,Zamolodchikov:1995aa} holomorphic and anti-holomorphic BPZ conformal blocks. The pairing is defined on the physical locus for the external Liouville momenta, $\alpha = \frac{Q}{2} +i\mb R$, and can be analytically continued to the complex $\alpha$ plane by setting the parameter in the anti-holomorphic conformal block to be $\bar \alpha = Q - \alpha$ (or $\bar \alpha = \alpha$, the choice is immaterial in BPZ conformal blocks). The collision limit is done by sending the Liouville momenta $\alpha_k$ of the colliding puncture to $\pm i \infty$, and involves 
the cancellations of divergent factors between the DOZZ three-point functions and the conformal blocks. 

We expect that an analogous collision limit should be possible in a suitable RCFT based on the $\wh{sl}(2)$ current algebra: the so-called $H_3^+$ WZW model. 
A standard primary field $\Phi_j(\mu, \bar \mu|z, \bar z)$ in this RCFT carries a spin $j\in-\tfrac12+i\mb R$. The correlation functions are defined on this physical slice of parameter space (and $\mu^* = \bar \mu$),
and can be analytically continued away from there, by keeping the spin in the anti-holomorphic informal blocks $\bar j = j$ (the convention $\bar j = -j-1$ is equivalent, but less convenient)\cite{Ribault:2005wp}. 

The collision limit gives a generalization of the physical slice $j\in-\tfrac12+i\mb R$. For a vector of rank $r$, we require $j_0\in-\tfrac{r+1}2+i\mb R$, and $\bar j_n=j_n=-j_n^*-(r+1)\delta_{n,0}$. 
\subsection{An example: one-point function of a rank two irregular vector}
\label{sec:one-point}

To show that the above results are consistent, we proceed with an example and calculate the one-point function of a rank two irregular vector, up to a function of $j_0$. This is the simplest nontrivial case, as rank 1 vectors have a vanishing one-point function by symmetry. We use the fact that the singular modes of the currents $T(w)$ and $J^a(w)$ annihilate the vacuum state. This implies that the differential operators $\mc L_0$, $\mc L_{\pm 1}$ and $\mc J_0^a$ must annihilate the one-point function. In the current case these conditions are sufficient to fix the correlation function, and the generalized KZ equations are not needed. The six equations read
\begin{align}
\mc J^-_{0}\la\Phi^{(2)}_{\bf j}({\boldsymbol \mu}|z)\ra=&\mu_0\la\Phi^{(2)}_{\bf j}({\boldsymbol \mu}|z)\ra=0,\nn
\mc J^0_{0}\la\Phi^{(2)}_{\bf j}({\boldsymbol \mu}|z)\ra=&(j_0-\mu_0\partial_{\mu_0}-\mu_1\partial_{\mu_1}-\mu_2\partial_{\mu_2})\la\Phi^{(2)}_{\bf j}({\boldsymbol \mu}|z)\ra=0,\nn
\mc J^+_{0}\la\Phi^{(2)}_{\bf j}({\boldsymbol \mu}|z)\ra=&(2j_0\partial_{\mu_0}+2j_1\partial_{\mu_1}+2j_2\partial_{\mu_2}\nn
&\quad-\mu_0\partial_{\mu_0}^2-2\mu_1\partial_{\mu_0}\partial_{\mu_1}-\mu_2(\partial_{\mu_1}^2+\partial_{\mu_0}\partial_{\mu_2})
)\la\Phi^{(2)}_{\bf j}({\boldsymbol \mu}|z)\ra=0,\nn
\mc L_{-1}\la\Phi^{(2)}_{\bf j}({\boldsymbol \mu}|z)\ra=&\partial_z\la\Phi^{(2)}_{\bf j}({\boldsymbol \mu}|z)\ra=0,\nn
\mc L_{0}\la\Phi^{(2)}_{\bf j}({\boldsymbol \mu}|z)\ra=&
(\mu_1\partial_{\mu_1}+2\mu_2\partial_{\mu_2}+j_1\partial_{j_1}+2j_2\partial_{j_2}
-b^2j_0(j_0+1))\la\Phi^{(2)}_{\bf j}({\boldsymbol \mu}|z)\ra=0,\nn
\mc L_{1}\la\Phi^{(2)}_{\bf j}({\boldsymbol \mu}|z)\ra=&
(\mu_2\partial_{\mu_1}+j_2\partial_{j_1}
-2b^2j_1(j_0+1))\la\Phi^{(2)}_{\bf j}({\boldsymbol \mu}|z)\ra=0.
\end{align}
This system, together with its anti-holomorphic counterpart, has the solution
\begin{align}
\la\Phi^{(2)}_{\bf j}({\boldsymbol \mu}|z)\ra=&
\delta^{(2)}(\mu_0)|\mu_2|^{2j_0+2}|j_2|^{(j_0+1)(b^2j_0-2)}e^{2ib^2(j_0+1)\text{Im}\l(\frac{j_1^2}{j_2}\r)+\frac12i\text{Im}(u^2)}\nn
&\times\l(
D_{j_0+1}(u),~
D_{-j_0-2}(iu)\r)C(j_0)
\l(D_{j_0+1}(\bar u),~
D_{-j_0-2}(i\bar u)\r)^\text{T},
\end{align}
where $u=i (\mu_2/2)^{-1/2}(j_2(\mu_1/\mu_2)-j_1)$, with $\bar u=-i u^*$, $C(j_0)$ is an arbitrary $2\times2$ matrix, and $D_A$ is the parabolic cylinder function 
\begin{align}
D_A(z)=2^{A/2}\sqrt{\pi}e^{-\frac14z^2}
\l(\Gamma\l(\tfrac{1-A}2\r)^{-1}\F{1}{1}\l(-\frac A2;\frac12;\frac{z^2}2\r)-\sqrt{2}~\Gamma\l(-\tfrac A2\r)^{-1}z
\F{1}{1}\l(\frac{1-A}2;\frac32;\frac{z^2}2\r)
\r).
\end{align}
To find the matrix $C(j_0)$ we impose the constraint that the physical correlation function do not grow exponentially in any direction in the $u$-plane. This property holds for the three-point function whose collision limit defines the correlation function we are after. %We will verify momentarily that the same  $C(j_0)$ indeed arises in the collision limit. \cb{This is quite far actually}
The parabolic cylinder function has a Stokes phenomenon, so we must make sure the constraint holds for each of the four Stokes sector\footnote{For a review of the Stokes phenomenon, see \cite{Witten:2007td}}. This fixes the matrix to $C(j_0)= $diag$(1,$ $i ~\gamma(j_0+2))$, where $\gamma(A)=\Gamma(A)/\Gamma(1-A)$, up to an overall factor.

The correlation function can be expressed as the double integral
\begin{align}\label{one-pt integral}
\la\Phi^{(2)}_{\bf j}({\boldsymbol \mu}|z)\ra= & 
\delta^{(2)}(\mu_0)C(j_0)|\mu_2|^{2j_0+2}|j_2|^{(j_0+1)(b^2j_0-2)}e^{2ib^2(j_0+1)\text{Im}\l(\frac{j_1^2}{j_2}\r)}\nn
&\times\int_{\mb C} dx d\bar x~|x|^{-2(j_0+2)}e^{-\frac12x^2+\frac12\bar x^2-ux+u^* \bar x}.
\end{align}
The integral is calculated in appendix \ref{app:parabolic cylinder}
The integrand is the ``square'' of the integrand in the contour integral definition of the parabolic cylinder function. The holomorphic and anti-holomorphic parameters are such that the integrand is oscillatory, of modulo $1$, and does not blow up exponentially at large $u$. The double integral can also be interpreted as the Fourier transform of the correlation function in the $x$-basis, which can be determined 
independently by the collision limit on the standard three-point function of the WZW model in the $x$-basis. In appendix \ref{app:one-point} we verify that the same correlation function (\ref{one-pt integral}) arrises from the collision limit (in the $\mu$-basis).

\section{Irregular vectors and the Liouville-H$_3^+$ duality}
\label{sec:duality}
\subsection{The Liouville-H$_3^+$ duality}

We review the duality between WZW and Liouville correlation functions. Primary fields in Liouville theory are parametrized by a Liouville momentum $\alpha$, and have conformal dimension $\Delta_\alpha=\alpha(Q-\alpha)$. Here $Q=b+b^{-1}$ is a parameter of the theory, related to the central charge by $c=1+6Q^2$. For a review of Liouville theory, see \cite{Harlow:2011ny,Teschner:2001rv}. 

The Liouville-H$_3^+$ duality relates the correlation functions of primary fields in the form \cite{Ribault:2005wp}
\begin{align}\label{duality}
\l\la\prod_k^n\Phi_{j^k}(\mu^k|z_k)\r\ra
=\frac\pi2(-\pi)^{-n}b~\delta^{(2)}(\sum_k^n\mu_k)~|\Theta_{n,n-2}|^2\l\la
(\prod_k^n V_{\alpha_k}(z_k))(\prod_m^{n-2} V_{-\frac{1}{2b}}(y_m))\r\ra.
\end{align}
where the function $\Theta_{n,m}$ is defined as\footnote{Here the function differs from the one defined in \cite{Ribault:2005wp} because of the different normalization for the fields.}
\begin{align}\label{dualitytheta}
\Theta_{n,m}(z_1,\cdots,z_n|y_1,\cdots,y_{m},u)=u^{\sum_{k}^n\frac{\alpha_k}{b}-\frac{n}{2b^2}-(n-1)}
\frac{\prod_{s,k<s}^n(z_k-z_s)^{-\frac{\alpha_k+\alpha_s}{b}+\frac{3}{2b^2}+2}}{\prod_{k}^n\prod_p^{m}(z_k-y_p)^{-\frac{\alpha_k}{b}+\frac1{b^2}+1}}
\prod_{p,q<p}^{m}(y_q-y_p)^{\frac{1}{2b^2}}.
\end{align}
The WZW variables are related to those on the Liouville side by
\begin{align}
\label{dualityalpha}
\alpha_k&=b(j^k+1)+\frac{1}{2b},\\
\label{dualitymu}
\mu^k&=u\frac{\prod_m^{n-2}(z_k-y_m)}{\prod_{s\neq k}^n(z_k-z_s)}.
\end{align}
The parameters $k$ and $b$ are related under the duality by $b^2=-(k+2)^{-1}$, and $u=\sum_{k=1}^n\mu^kz_k$.

This somewhat imposing relation can be understood more easily after two simple observations. 
First, the $y_m$ are essentially the zeroes of $J^-(y)$, or more precisely of 
\begin{equation}
\sum_k \frac{\mu^k}{y-z_k}.
\end{equation} This statement inverts the relation (\ref{dualitymu}). 
Second, the $\Theta_{n,m}$ prefactor, which allows one to reduce the KZ equations on the WZW side to the BPZ equations 
satisfied by the degenerate punctures at $y_m$, can be derived by comparing the various OPE limits of corresponding conformal blocks 
on the two sides. 
A similar relation holds at the level of conformal blocks. Each trinion on the WZW side maps to a trinion on the Liouville side with an
extra degenerate on one of the three legs. There is a two-dimensional space of three-point junctions on the WZW side, 
which matches the two-dimensional space of Virasoro three-point junctions with an extra degenerate insertion on one of the three legs. 
 
\subsection{$H_3^+$ irregular vectors from Liouville theory}
\label{sect:dualirreg}

We now turn to the main objective of this section, and look for an equivalent of eq. (\ref{duality}) for correlation functions involving irregular vectors. Again, we use the collision limit for this purpose. The duality formula provides an alternative definition of $H_3^+$ correlation functions involving irregular vectors, in terms of the corresponding Liouville theory correlation functions, which behave as described in section \ref{sec:Virasoro}. Most calculations in this section are tedious and are done in the appendix \ref{app:dualityIrreg}.

We look for the duality involving $n$ $H_3^+$ vectors of ranks $r_k$, which can be regular if $r_k=0$. For this purpose we start from the duality formula with $t=\sum_{k=1}^n(r_k+1)$ $H_3^+$ regular vectors. This implies on the Liouville side the presence of $t$ regular vectors and $t-2$ degenerate insertions. Then, we do the usual collision limit on the $H_3^+$ side. 
Because, by definition, we tune the $\mu^k$ so that the $J^-(y)$ Ward identity finite in the limit, the location of the zeroes $y_m$ remains generic in the collision limit. On the other hand, the way we tune the $j^k$ to keep the $J^0(y)$ Ward identity finite 
 is compatible with the standard collision limit on the Virasoro side of the duality. Thus the endpoint of the collision limit on the Liouville side 
involves irregular vectors of rank $r_k$ and $t-2$ degenerate fields. 

Thus our final formula includes $n$ fields $\Phi^{(r_k)}_{{\bf j}^k}({\boldsymbol \mu}^k|z_k)$ on the $H_3^+$ side and the dual fields $V^{(r_k)}_{{\bf c}^k}(z_k)$ and $t-2$ degenerate fields $V_{-\frac{1}{2b}}(y_p)$. The parameters are related by 
\begin{align}\label{dualityalphak}
c_{m}^k&=b j_{m}^k+(r+1)(b+\frac1{2b})\delta_{m,0},\\
\label{dualitymuk}
\mu_{m}^k&=u\frac{\prod_{p=1}^{t-2}(z_k-y_p)}{\prod_{s\neq k}^n(z_k-z_s)^{r_s+1}}W_{r_k-m}^k,
\end{align}
where $W_{r_k-m}^k$ represents the series expansion
\begin{align}\label{dualityWk}
\sum_pW_p^k q_i^p=\frac{\prod_m^{t-2}(1+q_i(z_k-y_m)^{-1})}{\prod_{s=r+2}^n(1+q_i(z_k-z_s)^{-1})^{r_s+1}}.
\end{align}
The first relation follows straightforwardly from (\ref{dualityalpha}), while the second requires more work and is derived in appendix \ref{app:dualityIrreg}. Of course, this relation is simply a complicated way to say that the $y_m$ are zeroes of the $J^-(y)$ Ward identity.

The duality formula for this field content is also derived in appendix \ref{app:dualityIrreg}, and takes the form
\begin{align}\label{dualityirreg}
&%\prod_{k=1}^n\l(|(A_{r+1})_k|^{2(B_1)_k+2r_k+2}\exp\l(
%\sum_{p=1}^{r}\frac{2}{p}\l|\frac{(A_{r})_k}{(A_{r+1})_k}\r|^p (B_{p+1})_k\r)\r)
\l\la \prod_{k=1}^n\Phi^{(r_k)}_{{\bf j}^k}({{\boldsymbol \mu}^k}|z_k)\r\ra=\frac\pi2(-\pi)^{-t}b~\delta^{(2)}(\sum_{k=1}^n\mu_0^k)
~|\Theta_{{\bf r}}|^2
\l\la (\prod_{k=1}^{n} V^{(r_k)}_{{\bf c}^k}(z_k))(\prod_m^{t-2} V_{-\frac{1}{2b}}(y_m))\r\ra.
\end{align}
The function $\Theta_{{\bf r}}$ is the generalization of $\Theta_{n,n-2}$ to irregular vectors and is given by
\begin{align}
&\Theta_{{\bf r}}(z_1,\cdots,z_n|y_1,\cdots,y_{t-2},u)
\nonumber\\
&\qquad=
	u^{\sum_k^n\frac{\alpha_k}{b}-\frac{t}{2b^2}-(t-1)}
	\prod_{p,q<p}^{t-2}(y_q-y_p)^{\frac{1}{2b^2}}
	\frac{\prod_{k,s<k}^{n}(z_k-z_s)^{-\frac{(r_s+1)\alpha_k+(r_k+1)\alpha_s}{b}+(r_k+1)(r_s+1)(2+\frac{3}{2b^2})}}
	{\prod_k^n\prod_m^{t-2}(z_k-y_m)^{-\frac{\alpha_k}{b}+(r+1)(1+\frac{1}{2b^2})}}
\nonumber\\
&\qquad\qquad\times\quad
	\frac{\prod_{k,s<k}^{n}
	\exp\l(b^{-1}\sum_{p=1}^{\max(r_k,r_s)}(-1)^{p}\frac{(r_s+1)c_{p}^k+(r_k+1)c_{p}^s}{p(z_k-z_s)^{p}}\r)}
	{\prod_k^n\prod_m^{t-2}
	\exp\l(b^{-1}\sum_{p=1}^{r}\frac{(-1)^{p}c_{p}^k}{p(z_k-y_m)^{p}}\r)}
	F^{(r)}({\boldsymbol j})
\end{align}
with $u=\sum_{k=1}^n(j_0^k z_k+j_1^k)$. The function $F^{(r)}({\boldsymbol j})$ is the exponent of a rational function of the ${\boldsymbol j}$, which is defined as 
the solution to the set of differential equations
\begin{align}\label{Fdef}
\sum_{n=0}^{r-m}nj_{m+n}\frac{\partial_{j_n}F^{(r)}({\boldsymbol j})}{F^{(r)}({\boldsymbol j})}
+2(r-m)(b^2+1)j_m+(b^2+1)r(r+1)\delta_{m0}=0,
\end{align}
for $0\leq m\leq r$, normalized such that $F(0,\cdots,0,j_r)=j_r^{(b^2+1)(r+1)}$. 
A similar formula holds at the level of conformal blocks. Because of the structure of the collision limit, we expect each of the irregular junctions
described in \cite{Gaiotto:2012sf} on the Virasoro side to combine with a degenerate insertion to give an irregular junction in irregular WZW conformal blocks. 
We leave the details of the dictionary between irregular conformal blocks to future work. 

\section{Semiclassical analysis of irregular vectors}
\label{sec:semiclassical}

In order to understand better the meaning of irregular vectors in WZW models, we can look at the semiclassical limit of correlation functions. 
The semiclassical limit is defined as a $k\to \infty$ limit, combined with an appropriate rescaling of the 
parameters at the punctures, in such a way that correlation functions scale as
\begin{equation}
\langle \prod_i \Phi_i(z_i) \rangle \sim e^{\frac{1}{k+2} S_{\mathrm{cl}}(z_i)} 
\end{equation}
in terms of the classical action for an appropriate solution of the classical WZW equations of motion, determined by the data at the punctures. 

Here the action is the WZW action for the gauge group $H_3^+=SL(2,C)/SU(2)$. One useful parametrization for a group element is 
\begin{align}\label{hparam}
h=\l(\begin{matrix}1&\gamma\\0&1\end{matrix}\r)
\l(\begin{matrix}e^{\phi}&0\\0&e^{\phi}\end{matrix}\r)
\l(\begin{matrix}1&0\\\bar\gamma&1\end{matrix}\r)
=\l(\begin{matrix}e^{-\phi}+\gamma\bar\gamma e^{\phi}&e^{\phi}\gamma\\e^{\phi}\bar\gamma&e^{\phi}\end{matrix}\r).
\end{align}
In terms of these parameters, the action reads
\begin{align}
S_{\text{WZW}}=\frac{k}{\pi}\int d^2z \text{Tr}(\partial h^{-1}\bar\partial h)+k\Gamma_{\text{WZ}}
=\frac{k}{\pi}\int d^2z(\partial\phi\bar\partial\phi+e^{2\phi}\partial\bar\gamma\bar\partial\gamma).
\end{align}
This also corresponds to a string action in $H_3^+$, with the metric 
$ds^2=d\phi^2+e^{2\phi}d\gamma d\bar\gamma.$. The $H_3^+$ theory is reviewed in \cite{deBoer:1998pp,
Maldacena:2000hw,Maldacena:2001km,Teschner:1997ft}

The KZ equations for a spin $1/2$ degenerate field is also expected to have a finite limit, and reduce to the equations of motion for the 
classical WZW solution. More precisely, the KZ differential operator should go to an ordinary differential operator in the semiclassical limit 
\begin{equation}
\partial_z - \frac{1}{k+2} J^a \sigma_a \to \partial_z - A(z).
\end{equation}
At regular singularities, $A(z)$ should have a single pole with residue $R=R^a\sigma_a$ 
\begin{align}
R^-=m,\qquad R^0=a-x m,\qquad R^+=2a x-x^2 m.
\end{align}
Here we scaled $j = a (k+2)$, $\mu = m (k+2)$, and $x = \frac{\partial S_{\mathrm{cl}}}{\partial \mu}$. 
This is a rather generic parameterization for a 
traceless matrix of fixed eigenvalues $a$ and $-a$ in terms of a pair of conjugate variables $x$ and $m$. 

The group element $G(z, \bar z)$ which represents a classical solution of the WZW equations of motion 
should satisfy
\begin{equation}
\partial G = A G, \qquad \bar \partial G = G \bar A.
\end{equation}
and it should be single-valued on the Riemann surface. 
The latter constraint is rather strong. We can solve the equations by writing 
\begin{equation}
G(z,\bar z) = g(z) C \bar g(\bar z).
\end{equation}
for some holomorphic solution $g(z)$ and constant hermitian matrix $C$. 

The holomorphic solution $g(z)$ will have monodromy 
\begin{equation}
g(e^{2 \pi i} (z-z_s) + z_s) = g(z) M_s
\end{equation}
around the regular singularities at the locations $z_s$ of the ordinary vertex operators. 
Then $G$ will be single-valued if 
\begin{equation} \label{eq:conj}
M_s C \bar M_s = C,
\end{equation}
i.e. $M_a$ is conjugate to $\bar M_s^{-1}$. 

This means that the trace of the monodromy along any path is real. 
This constraint kills half of the degrees of freedom of the system. In principle it fixes, say, 
the $m$ parameters at all punctures in terms of the $x$ parameters at all punctures. 
As the relation between $A$ and its monodromies is highly transcendental, 
it is very hard to describe the constraints on $A$ implied by the constraints on the monodromy matrices $M_s$. 
The semiclassical limit of WZW conformal blocks solves this problem. 
This is analogous to the statement that the semiclassical limit of Virasoro conformal blocks 
solves the uniformization problem. 

This constraints can only be satisfied if the parameter $a_s$ for the regular singularity at $z_s$ is either pure real or pure imaginary:
in the first case the eigenvalue $e^{2 \pi i a}$ of $M$ is a phase, in the second it is real. 
If we pick the $a_s$ parameters to be real at all punctures, then we can pick a gauge where all 
the monodromies are unitary matrices, and $C=1$. Then $G$ lives in the space of hermitean matrices of unit determinant, 
which is the same as the hyperbolic space $H_3^+$. This is the expected target space for the $H_3^+$ WZW sigma model. 

On the other hand, the vertex operators which represent normalizable states in the $H_3^+$ model have 
pure imaginary $a_s$. If we pick the $a_s$ parameters to be imaginary at all punctures, then we can pick a gauge where all 
the monodromies are real, i.e. belong to $SL(2,R)$. Then we need to pick $C = i \sigma_2$. 
As a consequence, $G$ lives in the space of hermitean matrices of determinant $-1$, 
which is an analytic continuation of $H_3^+$: it is three-dimensional de-Sitter space $dS_3$. 
It may seem strange for the semicalssical saddle points for correlation functions 
of normalizable vertex operators to take value in the complexification of the target space, 
but it becomes less surprising if we look at a much simpler CFT: the theory of a free boson. 
The semiclassical solutions near a normalizable vertex operator $e^{i p X}$ are imaginary
\begin{equation}
X \sim i p \alpha' \log |z|^2.
\end{equation}

This is also the correct range of parameters for the regular punctures we collide to obtain irregular punctures. 
Our expectation is based on the Liouville theory analogue setup \cite{Gaiotto:2012sf}. It essentially means that in the collision limit, 
one of the eigenvalues of each of the monodromy matrices $M_s$ involved in the collision should be sent to infinity.  
We will see momentarily that the natural constraints at irregular singularities which replace the single-valuedness at regular singularities 
indeed require a $C$ of the form $C = i \sigma_2$.

In order to understand the semiclassical behaviour near a regular puncture, 
it is useful to consider solutions $g^{(s)}_\pm(z)$ which behave as 
$(z-z_s)^{\pm a}$ at the regular singularity. If $a_s$ is real, then the solution must take the diagonal form
\begin{equation}
G = c_s g^{(s)}_+ \left(g^{(s)}_+\right)^* + c_s^{-1} g^{(s)}_- \left(g^{(s)}_-\right)^*.
\end{equation}
The coefficient $c_s$ is fixed by the requirement that $G$ should take the diagonal form at all punctures. 
As we approach the singularity, one of the two solutions blows up, and $G$ goes to the boundary of $H_3^+$,
at a location determined by the $x$ parameter of the regular puncture
\begin{equation}
G \sim \begin{pmatrix}
1 & \bar x \cr x & x \bar x 
\end{pmatrix} |z-z_s|^{- 2 |a_s|}.
\end{equation}
Thus the semiclassical solution is the solution of the equation of motion which reach the boundary at a prescribed set of points. 

If $a_s$ is imaginary, then the solution must take the off-diagonal form
\begin{equation}
G = c_s g^{(s)}_+ \left(g^{(s)}_-\right)^*+ c_s^{-1} g^{(s)}_- \left(g^{(s)}_+\right)^*.
\end{equation}
As we approach the singularity, neither solution diverges. Rather, we get an oscillating approximate solution 
\begin{equation} G \to \begin{pmatrix}
1 \cr x
\end{pmatrix} \begin{pmatrix}
\bar m & 2 \bar a + \bar m \bar x \end{pmatrix} |z-z_s|^{2 i |a_s|} + \begin{pmatrix}
m \cr 2 a + m x
\end{pmatrix} \begin{pmatrix}
1 &  \bar x \end{pmatrix} |z-z_s|^{-2 i |a_s|}.
\end{equation}
As we approach the regular singularity, the solution winds infinitely many times along 
a specific circle in $dS_3$.

At an irregular vertex operator, the differential operator $\partial_z - A$ has an irregular singularity:
the matrix $A$ takes the form $A=\sum_{n=0}z^{-n-1}R_n^a\sigma_a$, where
\begin{align}
R^-_n=m_n,\qquad R^0_n=a_n-\sum_p m_n x_{p-n},\qquad R^+_n=2\sum_p a_n x_{p-n}-\sum_{p,q}m_px_qx_{p-q-n+1}.
\end{align}
To simplify the analysis, we set $x_0=m_r=0$ by a $H_3^+$ transfromation, and $z_s=0$.
The solution $g(z)$ will have Stokes phenomena. Given a generic straight ray going into the irregular singularity, we can find a unique solution which decreases exponentially fast along that ray according to a specific asymptotic behavior, 
which is valid only in an appropriate Stokes sector around the ray.
Roughly, 
\begin{align}\label{AsympHolomorphic}
g(z)\sim z^{\pm a_0}e^{\mp \sum_{n=1}^r\frac{a_n}{nz^n}}\end{align}
with an appropriate vector structure. 
This procedure identifies $2n$ ``small solutions'' 
$g^{(s)}_i$, each decreasing exponentially fast in a sector of width $\pi/n$ around the irregular singularity. 

Pairs of consecutive small solutions $g^{(s)}_i$, $g^{(s)}_{i+1}$ are always linearly independent, and can be normalized so that $\det (g^{(s)}_i,g^{(s)}_{i+1})=1$. If we compare $g^{(s)}_{i+1}$ and $g^{(s)}_{i-1}$
in the $i$-th sector, they will grow at the same rate, and their sum will be proportional to $g^{(s)}_i$:
\begin{equation} \label{eq:stokes}
g^{(s)}_{i+1}+g^{(s)}_{i-1} = d^{(s)}_i g^{(s)}_i.
\end{equation}
Imposing this normalization, we will get a periodicity $g^{(s)}_{i+2n} = \eta^{\pm1} g^{(s)}_{i}$, where the ``formal monodromy''
$\eta$ depends on $a_0$. 

The proportionality coefficients $d^{(s)}_i$ generalize the notion of monodromy around a regular puncture.
In particular, if we look at an irregular singularity as the collision of $n$ regular singularities, 
the coefficients $d^{(s)}_i$ control the behaviour of the $n$ monodromy matrices in the limit. 

We should ask what is the condition on the $d^{(s)}_i$ which arises from the condition (\ref{eq:conj}) in the collision limit. 
We can take a shortcut: as the solution $G$ did not blow up at the regular singularities, it should also not blow up at an irregular singularity. 
Thus between each pair of sectors $G$ should take the off-diagonal form proportional to 
\begin{equation}
g^{(s)}_i \left(g^{(s)}_{i-1} \right)^*- g^{(s)}_{i-1} \left(g^{(s)}_i\right)^*.
\end{equation}
This is compatible with (\ref{eq:stokes}) if the $d^{(s)}_i$ are real. The asymptotic solution in all sectors is, to leading order, of the form
\begin{align}
G(z,\bar z)\sim& \l(\begin{array}{cc}
-2\text{ Re}(x_1zS(z,\bar z))&S^*(z,\bar z)%-\frac{x_1\bar m_{r-1}}{2\bar a_r}|z|^2S(z,\bar z)
\\
S(z,\bar z)%-\frac{\bar x_1 m_{r-1}}{2 a_r}|z|^2S^*(z,\bar z)
&2\text{ Re}\l(\frac{m_{r-1}}{2 a_r}zS^*(z,\bar z)\r)
\end{array}\r)+\mc O(z^2),
\end{align}
where we defined
\begin{align}
S(z,\bar z)=|z|^{2a_0}e^{-2i\text{Im}\sum_{n=1}\frac{a_n}{nz^n}}=\exp\l[2i\text{Im}\l(a_0\log |z|-\sum_{n=1}\frac{a_n}{nz^n}\r)\r]=e^{i\psi}.
\end{align}
At the zeroth order (neglecting the $\mc O(z)$ terms), the solution is included in a U(1) subgroup of dS$_3$ parametrized by the angle $\psi$. For sufficiently small and constant $|z|$ the solution winds quickly in alternating direction, i.e., $\psi'(\theta)$ is separated in $2n$ sectors of opposing sign. 

\acknowledgments{The research of DG and JLP was supported by the Perimeter Institute for Theoretical
Physics.  Research at Perimeter Institute is supported by the
Government of Canada through Industry Canada and by the Province of
Ontario through the Ministry of Economic Development and Innovation.}

\begin{appendix}

\section{Collision limits}
\label{app:coll}

\subsection{The irregular Virasoro current}
\label{app:sl2Vir}

Here we derive the irregular Virasoro current (\ref{VirasoroWard}) for $\wh{sl}(2)$ WZW models from a collision limit. As in section \ref{sec:sl2} we start with $r+1$ primary fields $\Phi_{j^k}(\mu^k|z+q_k)$, and write the collision in the form
\begin{align}
&\Phi'^{(r)}_{\bf j}({\boldsymbol \mu}|z,{\bf q})=\prod_{i,j<i}|q_i-q_j|^{2b^2C_{ij}}\prod_{k=1}^{r+1}\Phi_{j^k}(\mu^k|z+q_k),\nn
&\text{with}\quad \tilde\Phi^{(r)}_{\bf j}({\boldsymbol \mu}|z)=\lim_{{\bf q}\rw{\bf 0}}\Phi'^{(r)}_{\bf j}({\boldsymbol \mu}|z,{\bf q}).
\end{align}
As will be shown in the next section, finiteness of the conformal blocks in the collision limite require $C_{ij}=2j_ij_j+2(1+b^{-2})(j_i+j_j+1)$. We look for the collision limit of the Virasoro Ward identity
\begin{align}
T(w)\prod_{k=1}^{r+1}\Phi_{j^k}(\mu^k|z+q_k)
\sim&\sum_{k=1}^{r+1}\l(\frac{\Delta_k}{(w-z-q_k)^2}+\frac{\partial_{z_k}}{(w-z-q_k)}\r)\nonumber\\
&\qquad\times\prod_{i,j<i}|q_i-q_j|^{-2b^2C_{ij}}\Phi'^{(r)}_{\bf j}({\boldsymbol \mu}|z,{\bf q}),
\end{align}
with $\Delta_k=-b^2j^k(j^k+1)$. Defining $C_{ji}=C_{ij}$ for $j>i$, this becomes
\begin{align}%\label{VirasoroWard}
T(w)\Phi'^{(r)}_{\bf j}({\boldsymbol \mu}|z,{\bf q})
\sim&\l(\sum_{m=0}^{\infty}\frac{-(m+1)b^2\sum_i q_i^{m}j^i (j^i+1)-\sum_{i,j\neq i}^{r+1}C_{ij}b^2q_i^{m+1}(q_i-q_j)^{-1}
}{(w-z)^{m+2}}\r.\nonumber\\
&\l.+\sum_{m=-1}^{\infty}\frac{\sum_{i}^{r+1} q_i^{m+1}\partial_{z_i}}{(w-z)^{m+2}}\r)\Phi'^{(r)}_{\bf j}({\boldsymbol \mu}|z,{\bf q}).
\end{align}
For the moment we neglect the part with the derivative terms, and call the rest $T_j(w)$. The terms containing $C_{ij}$ can berewritten in the form
\begin{align}
\sum_{i,j\neq i}^{r+1}C_{ij}q_i^{m+1}(q_i-q_j)^{-1}
=&\frac12\sum_{i,j\neq i}^{r+1}\sum_{n=0}^{m}q_i^{n} q_j^{m-n-1}C_{ij},
\end{align}
which implies
\begin{align}
&T_j(w)\Phi'^{(r)}_{\bf j}({\boldsymbol \mu}|z,{\bf q})\nn
&\quad\sim-b^2\sum_{m=0}^{\infty}\frac{(m+1)(j_{m}+\sum_i q_i^{m}(j^i)^2)
+\sum_{i,j\neq i}^{r+1}\sum_{n=0}^{m}q_i^{n} q_j^{m-n-1}C_{ij}/2
}{(w-z)^{m+2}}\Phi'^{(r)}_{\bf j}({\boldsymbol \mu}|z,{\bf q})\nn
&\quad\sim-\sum_{m=0}^{\infty}\l(\frac{(b^2(2r+1-m)+2(r-m))j_{m}
}{(w-z)^{+2}m}\r.\nn
&\qquad\qquad\l.+
\frac{b^2\sum_{n=0}^{m}j_{n} j_{m-n-1}+(b^2+1)r(r+1)\delta_{m,0}}{(w-z)^{m+2}}
\r)
\Phi'^{(r)}_{\bf j}({\boldsymbol \mu}|z,{\bf q})
\end{align}

We now turn to the other part of the Virasoro Ward identity, containing $z$-derivatives (labeled $T_\partial(w)$). This part cannot easily be treated symmetrically in $i$, so here we set $q_{r+1}$ to 0. This gives the set of derivatives
\begin{align}
\partial_{z_{r+1}}=\partial_z-\sum_{i=1}^{r}\partial_{q_i},\qquad\text{and}\qquad
\partial_{z_i}=\partial_{q_i},\quad i\leq i\leq r.
\end{align}
We use this to rewrite
\begin{align}
T_\partial(w)\Phi'^{(r)}_{\bf j}({\boldsymbol \mu}|z,{\bf q})\sim\l(\frac{\partial_z}{(w-z)}+\sum_{m=0}^\infty\frac{\sum_{i}^{r} q_i^{m+1}\partial_{q_i}}{(w-z)^{m+2}}\r)\Phi'^{(r)}_{\bf j}({\boldsymbol \mu}|z,{\bf q}).
\end{align}
Using the chain rule to rewrite the $\partial_{q_i}$ as combinations of $\partial_{j_n}$ and $\partial_{\mu_n}$, we find
\begin{align}
T_\partial(w)\Phi'^{(r)}_{\bf j}({\boldsymbol \mu}|z,{\bf q})
\sim&\l(\frac{\partial_z}{(w-z)}+\sum_{m=0}^\infty\frac{\sum_{n=1}^{r+1}n(\mu_{n+m}\partial_{\mu_{n}}+j_{n+m}\partial_{j_{n}})}{(w-z)^{m+2}}\r)\Phi'^{(r)}_{\bf j}({\boldsymbol \mu}|z,{\bf q}).
\end{align}
Assembling the pieces together and taking the collision limit, we get the full Virasoro Ward identity (\ref{VirasoroWard})
\begin{align}
T(w)\tilde\Phi^{(r)}_{\bf j}({\boldsymbol \mu}|z)
\sim&
\l(-\sum_{m=0}^{2r}\frac{(b^2(2r+1-m)+2(r-m))j_{m}
}{(w-z)^{m+2}}\r.\nn
&\qquad-\sum_{m=0}^{2r}
\frac{b^2\sum_{n=0}^{m}j_n j_{m-n}+(b^2+1)r(r+1)\delta_{m,0}}{(w-z)^{m+2}}\nn
&\qquad+\l.\frac{\partial_z}{(w-z)}+\sum_{m=0}^{r-1}\frac{\sum_{n=1}^{r}n(\mu_{n+m}\partial_{\mu_n}+j_{n+m}\partial_{j_n})}{(w-z)^{m+2}}\r)\tilde\Phi^{(r)}_{\bf j}({\boldsymbol \mu}|z).
\end{align}
This differs from the expected result, but the difference is only due to a different scaling in $j_m$. We can obtain a Ward identity in the form (\ref{VirasoroWard}) with a rescaling
\begin{align}
\Phi^{(r)}_{\bf j}({\boldsymbol \mu}|z)=F^{(r)}({\boldsymbol j})\tilde\Phi^{(r)}_{\bf j}({\boldsymbol \mu}|z),
\end{align}
where $F^{(r)}({\boldsymbol j})$ is the function introduced in section \ref{sect:dualirreg}. By requiring $\Phi^{(r)}$ to satisfy (\ref{VirasoroWard}) we recover the defining set of differential equations (\ref{Fdef}).

\subsection{The duality formula for irregular vectors}
\label{app:dualityIrreg}

We evaluate the collision limit of eq (\ref{dualitymu}) to find the relation between the $\mu_m^k$ and the Liouville parameters in the duality. The setup is described in section \ref{sect:dualirreg}. Starting with the collisions which do not involve $\mu^i$, we find (assuming the field $i$ collide to the irregular field $k$)
\begin{align}
\mu^i&=u\frac{\prod_p^{t-2}(z_k+q_i-y_p)}{\prod_{j\neq i}^{r_k+1}(q_i-q_j)\prod_{m\neq k}^n(z_k+q_i-z_m)^{r_m+1}}.
\end{align}
For a regular field $r_k=0$ there is no factor $(q_i-q_j)$ in the denominator, and we can directly find the result by setting $q_i=0$, $\mu^i=\mu^k$. For the irregular case we expand in series in $q_i$
\begin{align}
\mu^i&=u\frac{\prod_p^{t-2}(z_k-y_p)}{\prod_{j\neq i}^{r_k+1}(q_i-q_j)\prod_{m\neq k}^n(z_k-z_m)^{r_m+1}}\sum_{p=0}^\infty q_i^p~W_p^k,
\end{align}
Where the $W_p^k$are the series coefficients defined in (\ref{dualityWk}). To compare with the parametrization found in section \ref{sec:sl2}, we calculate the sums $\sum_iq_i^x\mu^i$ explicitly (where the sum runs over the colliding fields)\footnote{To get the third line we can use the fact that the sum over $i$ is antisymmetric under odd permutations of the $q_k$, so it has to factor the Vandermonde determinant. For $x+p\neq r$, this fact fixes the sum to 0 or $\mc O({\bf q})$ times the denominator. For $x+p=r$ we can compare to the Laplace expansion of the Vandermonde determinant }:
\begin{align}
\sum_iq_i^{m}\mu^i
=&u\frac{\prod_p^{t-2}(z_k-y_p)}{\prod_{s\neq k}^n(z_k-z_s)^{r_s+1}}
\l(\sum_{p=0}^\infty W_p^k\sum_i\frac{q_i^{m+p}}{\prod_{j\neq i}^{r_k+1}(q_i-q_j)}\r)\nonumber\\
=&u\frac{\prod_p^{t-2}(z_k-y_p)}{\prod_{s\neq k}^n(z_k-z_s)^{r_s+1}}
\l(\sum_{p=0}^\infty W_p^k\frac{\sum_i(-1)^{r_k+1-i}q_i^{m+p}\prod_{s,t<s,s,t\neq i}(q_s-q_t)}{\prod_{s,t< s}(q_s-q_t)}\r)\nonumber\\
=&u\frac{\prod_p^{t-2}(z_k-y_p)}{\prod_{s\neq k}^n(z_k-z_s)^{r_s+1}}W_{r_k-m}+\mc O({\bf q}).
\end{align}
Taking the collision limit, we recover the announced result (\ref{dualitymuk})

We now turn to the limit of the full duality formula (\ref{duality}). Using the results of the previous section, we write
\begin{align}
&\prod_{k}\prod_{i,j<i}|q_{ik}-q_{jk}|^{-4b^4j^{ik}j^{jk}-4(b^2+1)(j^{ik}+j^{jk}+1)}|F({\boldsymbol j})|^2
\l\la \prod_{k}\Phi'^{(r_k)}_{{\bf j}^k}({\boldsymbol \mu}^k|z_k,{\bf q}_k)\r\ra\nonumber\\
&\qquad=\frac\pi2(-\pi)^{-t}b~\delta^{(2)}(\sum_k^n\mu^{ik})~|\Theta_{t,t-2}|^2\prod_{k}\prod_{i,j<i}|q_{ik}-q_{jk}|^{-4\alpha_{ik}\alpha_{jk}}
\l\la \prod_{k}V'^{(r_k)}_{{\boldsymbol \alpha}_k}(z_k,{\bf q}_k)\prod_p^{t-2} V_{-\frac{1}{2b}}(y_p)\r\ra,
\end{align}
where a pair of indices of the form $q_{ik}$ denotes the field number $i$ in the collision forming the irregular field $k$, for example ${\bf q}_k=(q_{1k},\cdots,q_{r_k+1~k})$. The function $\Theta_{t,t-2}$ goes as
\begin{align}
&F^{(r)}({\boldsymbol j})\Theta_{t,t-2}(z_1+{\bf q}_1,\cdots,z_n+{\bf q}_n|y_1,\cdots,y_{m},u)=u^{\sum_{ik}\frac{\alpha_{ik}}{b}-\frac{t}{2b^2}-(t-1)}\prod_{p,q<p}^{m}(y_q-y_p)^{\frac{1}{2b^2}}\nn
&\quad\times F^{(r)}({\boldsymbol j})\prod_{k}\prod_{i,j<i}(q_{ik}-q_{jk})^{-\frac{\alpha_{ik}+\alpha_{jk}}{b}+\frac{3}{2b^2}+2}
\frac{\prod_{s,k<s}\prod_{i,j}(z_{k}-z_s+q_{ik}-q_{js})^{-\frac{\alpha_{ik}+\alpha_{js}}{b}+\frac{3}{2b^2}+2}}{\prod_{ik}\prod_p(z_k+q_{ik}-y_p)^{-\frac{\alpha_{ik}}{b}+\frac1{b^2}+1}}
.\nn
&=u^{\sum_{k}\frac{c_{0}^k}{b}-\frac{t}{2b^2}-(t-1)}F^{(r)}({\boldsymbol j})
\prod_{k}\prod_{i,j<i}(q_{ik}-q_{jk})^{-\frac{\alpha_{ik}+\alpha_{jk}}{b}+\frac{3}{2b^2}+2}
\prod_{p,q<p}^{m}(y_q-y_p)^{\frac{1}{2b^2}}\nn
&\quad\times
\frac{\prod_{s,k<s}(z_{k}-z_s)^{-\frac{(r_s+1)c_0^k+(r_k+1)c_0^s}{b}+(r_k+1)(r_s+1)\l(\frac{3}{2b^2}+2\r)}
\exp\l(\sum_{p=0}^{\infty}
\frac{(r_k+1)c_p^s+(-1)^p(r_s+1)c_p^{k})}{bp(z_k-z_s)^{p}}\r)
}
{\prod_{k}\prod_p(z_k-y_p)^{-\frac{c_0^{k}}{b}+(r_k+1)\l(\frac1{b^2}+1\r)}
\exp\l(\sum_{p=0}^{\infty}\frac{c_p^k}{bp(y_p-z_k)^{p}}\r)}\nn
&\sim\prod_{k}\prod_{i,j<i}(q_{ik}-q_{jk})^{-\frac{\alpha_{ik}+\alpha_{jk}}{b}+\frac{3}{2b^2}+2}
\Theta_{{\bf r}}(z_1,\cdots,z_n|y_1,\cdots,y_{t-2},u).
\end{align}
The powers of $(q_{ik}-q_{jk})$ cancel with the ones already present in the duality formula. This is the announced validation for the choice of the rescaling for the irregular vector, as if the choice had been different some powers of the $(q_{ik}-q_{jk})$ would remain, leading to an uninteresting limit. Taking the collision limit, we recover the duality formula (\ref{dualityirreg}).

\section{The generalized KZ equations for irregular vectors}
\label{app:KZ}

In this appendix we provide the generalization of the KZ equation for $\wh{sl}(2)$ theories involving irregular vectors, as well as an outline of its derivation. Here we find the generalization from scratch in a way analogous to the  regular case \cite{CFT}, but it can also be found by taking the collision limit of the usual KZ equation. The starting point is the Sugawara construction for the energy-momentum operator in $\wh{sl}(2)$ WZW models:
\begin{align}\label{Sugawara}
T(w)=\frac{1}{2(k+2)}(J^aJ^a)(w).
\end{align}
The KZ equation is obtained by requiring consistency of both sides when inserted in correlation functions. This amounts to imposing the equality of the OPE for singular modes. For regular vectors, the positive modes annihilate the states, and the equation for the $L_0$ mode is trivially realized, so there is only one equation coming from $L_{-1}$. This equation reads \cite{CFT}
\begin{align}\label{KZregular}
\l(\partial_{z}+\frac{1}{k+2}\sum_{i=1}^n\frac{\mc D^a\mc D^a_i}{z-z_i}\r)\la\Phi_j(\mu|z)\Phi_{j_1}(\mu_1|z_1)\cdots\Phi_{j_n}(\mu_n|z_n)\ra=0.
\end{align}
For irregular vectors, some positive modes act nontrivially, so we have to generalize the consistency condition. The mode expansion of eq. (\ref{Sugawara}) reads
\begin{align}
L_n=&\frac{1}{2(k+2)}\sum_m :J_{m-n}^aJ_m^a:=\l\{\begin{array}{ll}\frac12J^a_{n/2}J^a_{n/2}+\sum_{m=1} J_{n/2-m}^aJ_{n/2+m}^a,& n\text{ even},\\
\sum_{m=0} J_{(n-1)/2-m}^aJ_{(n+1)/2+m}^a,& n\text{ odd}.
\end{array}\r.
\end{align}
We insert this equality inside a correlation function by applying it to a field $\Phi^{(r)}_{\bf B}({\bf A}|z)$, in the presence of $k$ other fields $\Phi^{(r_i)}_{\bf B_i}({\bf A_i}|z_i)$. This forces the equality
\begin{align}
\l\la\l(\l(L_n-\frac{1}{2(k+2)}\sum_m J_{m-n}^aJ_m^a\r)\Phi^{(r)}_{\bf B}({\bf A}|z)\r)(z)\prod_{i=1}^k\Phi^{(r_i)}_{\bf B_i}({\bf A_i}|z_i)\r\ra=0,\quad n\geq -1.
\end{align}
The effect of the irregular modes of $T(w)$ and $J^a(w)$ is known from eq. (\ref{IrregCurrents}) and (\ref{VirasoroWard}). However, there are also contributions from the non-singular part of $J^a(w)$. These can be expressed in terms of the other fields of the correlation function using the residue theorem ($n>0$):
%\begin{align}
%\l\la J_{-n}^a\Phi^{(r)}(z)\prod_{i=1}^k\Phi_{j_i}(\mu_i|z_i)\r\ra=0,
%\end{align}
%so that 
\begin{align}
&\l\la (J_{-n}^a\Phi^{(r)})(z)\prod_{i=1}^k\Phi_{j_i}^{(r_i)}(\mu_i|z_i)\r\ra= \oint\frac{dw}{(w-z)^n}\l\la (J^a\Phi^{(r)})(z)\prod_{i=1}^k\Phi_{j_i}^{(r_i)}(\mu_i|z_i)\r\ra
\nonumber\\
%&=-\sum_{s=1}^k\l\la\Phi^{(r)}(z)(J_{-n}^a\Phi_{j_s}^{(r_s)})(\mu_s|z_s)\prod_{i=1,i\neq s}^k\Phi_{j_i}^{(r_i)}(\mu_i|z_i)\r\ra\nonumber\\
&\qquad=-\sum_{s=1}^k \oint\frac{dw}{(w-z)^n}\l\la\Phi^{(r)}(z)(J^a\Phi_{j_s}^{(r_s)})(\mu_s|z_s)\prod_{i=1,i\neq s}^k\Phi_{j_i}^{(r_i)}(\mu_i|z_i)\r\ra\nonumber\\
%&\qquad=-\sum_{s=1}^k \oint\frac{dw}{(w-z)^n}\sum_{m=0}^{r_s}\frac{\mathcal J_{m(s)}^a}{(w-z_s)^{m+1}}
%\l\la\Phi^{(r)}(z)\prod_{i=1}^k\Phi_{j_i}^{(r_i)}(\mu_i|z_i)\r\ra\nonumber\\
%&\qquad=-\sum_{s=1}^k \sum_{m=0}^{r_s} \oint\frac{dw}{(w-z_s)^{m+1}}\frac{\mathcal J_{m(s)}^a}{(w-z)^n}
%\l\la\Phi^{(r)}(z)\prod_{i=1}^k\Phi_{j_i}^{(r_i)}(\mu_i|z_i)\r\ra\nonumber\\
%&\qquad=-\sum_{s=1}^k \sum_{m=0}^{r_s} \l[\l(\frac{\partial}{\partial w}\r)^m\frac{\mathcal J_{m(s)}^a}{(w-z)^n}\r]_{w=z_s}
%\l\la\Phi^{(r)}(z)\prod_{i=1}^k\Phi_{j_i}^{(r_i)}(\mu_i|z_i)\r\ra\nonumber\\
%&\qquad=-\sum_{s=1}^k \sum_{m=0}^{r_s}\frac{(-n)(-n-1)\cdots(-n-m+1)\mathcal J_{m(s)}^a}{(z_s-z)^{n+m}}
%\l\la\Phi^{(r)}(z)\prod_{i=1}^k\Phi_{j_i}^{(r_i)}(\mu_i|z_i)\r\ra\nonumber\\
&\qquad=-\sum_{s=1}^k \sum_{m=0}^{r_s}\frac{(-1)^m (n+m-1)!\mathcal J_{m(s)}^a}{(n-1)!(z_s-z)^{n+m}}
\l\la\Phi^{(r)}(z)\prod_{i=1}^k\Phi_{j_i}^{(r_i)}(\mu_i|z_i)\r\ra\nonumber\\
%&=\sum_{m=0}^{r_s}\frac{\mathcal J_{m(s)}^a}{(z-z_s)^{n+m}}\l\la\Phi^{(r)}(z)\prod_{i=1}^k\Phi_{j_i}^{(r_i)}(\mu_i|z_i)\r\ra,
\end{align}
where $\mathcal J_{m(s)}^a$ is the differential operator representing the action of $J_{m}^a$ on the field $s$. For the special case of a regular vector, we have only the term $m=0$, with $\mathcal J_{0(s)}^a=\mathcal D_s^a$. For even $n$ the equation becomes
\begin{align}
0=&\l(\mathcal L_{2n}-\frac{1}{k+2}\l(\frac12\mathcal J^a_n\mathcal J^a_n+\sum_{m=1}^{n} \mathcal J_{n-m}^a\mathcal J_{n+m}^a -\sum_{m=n+1}^{r-n} \sum_{s=1}^k\sum_{p=0}^{r_s}\frac{(-1)^p (m-n+p-1)!\mathcal J_{p(s)}^a \mathcal J_{n+m}^a}{(m-n-1)!(z_s-z)^{m-n+p}}\r)\r)\nonumber\\
&\qquad\times\l\la\Phi^{(r)}(z)\prod_{i=1}^k\Phi_{j_i}^{(r_i)}(\mu_i|z_i)\r\ra,\qquad n\geq 0.
\end{align}
Similarly, for odd $n$, 
\begin{align}
0=&\l(\mathcal L_{2n+1}-\frac{1}{k+2}\l(\sum_{m=1}^{n} \mathcal J_{n+1-m}^a\mathcal J_{n+m}^a -\sum_{m=n}^{r-n}\sum_{s=1}^k\sum_{p=0}^{r_s}\frac{(-1)^p (m-n+p-2)!\mathcal J_{p(s)}^a \mathcal J_{n+m}^a}{(m-n-2)!(z_s-z)^{m-n+1+p}}\r)\r)\nonumber\\
&\qquad\times\l\la\Phi^{(r)}(z)\prod_{i=1}^k\Phi_{j_i}^{(r_i)}(\mu_i|z_i)\r\ra,\qquad n\geq -1.
\end{align}
This is the generalized form of the KZ equation for $\wh{sl}(2)$ theories\footnote{Actually, this formula is completely general and holds for any sort of field in any WZW model (provided we replace the factor $(k+2)$ by $(k+g)$). What differs between theories is the actual form of the differential operators and the algebra in which the index $a$ is valued.}. The differential operators are given by (\ref{IrregCurrents}), (\ref{VirasoroWard}).

\section{Review of double integrals}
\label{app:integrals}

We discuss some integral identities, all of which are variations on the theme of the Riemann bilinear identity, 
which relates an integral over a Riemann surface to a bilinear of contour integrals over a basis of 1-cycles on the surface
\begin{equation}
\int_\Sigma \omega \wedge \tilde \omega = \sum_i \oint_{\alpha_i} \omega \oint_{\beta_i} \tilde \omega - \oint_{\beta_i} \omega \oint_{\alpha_i} \tilde \omega 
\end{equation}
Here $\omega$ is a holomorphic $(1,0)$ form, $\tilde \omega$ is an anti-holomorphic $(0,1)$ form, and the basis of cycles $\alpha_i$, $\beta_j$ 
is chosen as usual so that the intersection matrix is $\langle \alpha_i, \alpha_j \rangle = \langle \beta_i, \beta_j \rangle=0$, $\langle \alpha_i, \beta_j \rangle =\delta_{ij}$. 
The formula has obvious extensions to higher dimensional manifolds. 

A standard strategy to prove this relation is to look at this integral as a contour integral in $\Sigma \times \bar \Sigma$, 
with local coordinates $z$ and $\tilde z$ and integration contour ${\mathcal I}: \tilde z = \bar z$
Then we can simply decompose the integration contour into a basis for the second homology of  $\Sigma \times \bar \Sigma$,
which can be taken to consist of cycles of the form $\alpha_i \times \bar \alpha_j$, $\alpha_i \times \bar \beta_j$, etc. 
The coefficient of a basis element in the decomposition is simply the intersection number of ${\mathcal I}$ with a dual basis element. 
For example, the coefficient of $\alpha_i \times \bar \beta_j$ is the intersection number of ${\mathcal I}$ with $-\beta_i \times \bar \alpha_j$,
which is equal to the intersection of $\alpha_j$ and $\beta_i$ (some orientation sleight of hand here...), etcetera. 
This immediately leads to the bilinear identity.  
It is also useful to write the identity in terms of a generic basis of cycles $\gamma_a$ with intersection matrix $I_{ab}$: 
\begin{equation}
\int_\Sigma \omega \wedge \tilde \omega = - \oint_{\gamma_a} \omega I^{-1}_{ab} \oint_{\gamma_b} \tilde \omega 
\end{equation}

It is useful to give a trivial example of the bilinear identity. Consider the area of a torus, 
\begin{equation}
\int_{E_\tau} dz d\bar z = -2 i \mathrm{Im} \tau
\end{equation} 
The period of $dz$ on $\alpha$ is $1$, and on $\beta$ it is $\tau$. 

\subsection{Multivaluedness}
We will need two simple generalizations of this strategy. The first is to consider situations where $\omega \wedge \tilde \omega$ 
is single-valued, but $\omega$ is not. The second is to consider non-compact situations where $\omega$ may diverge at infinity, while $\omega \wedge \tilde \omega$ is integrable. 

Consider a situation where $\omega$ is a section of some line bundle, i.e. it is a multi-valued holomorphic $(1,0)$ form with constant Abelian monodromies 
$\omega \to \mu_p \omega$ when transported along some closed path $p$. We take the $\mu_p$ to be monomials in a certain set of $n$ generators $\mu_s$. 
Suppose that $\tilde \omega$ has opposite monodromies, 
so that $\omega \wedge \tilde \omega$ is single-valued, and the integral 
\begin{equation}
\int_\Sigma \omega \wedge \tilde \omega 
\end{equation}
can still be considered as a contour integral on ${\mathcal I}$. Now the contour integrals for $\omega$
do not belong to the homology of $\Sigma$. We can consider a cover $\Sigma_\mu$ of $\Sigma$ on which $\omega$ is single-valued, 
and work with the homology of that cover. We can take the cover to have fiber $\mathbb{Z}^n$, gluing it together by the map $p \to \mu_p$.

As we only really care about integration cycles for $\omega$,  we can naturally represent the images of a cycle $\gamma$ under $\mathbb{Z}^n$ deck transformations
as $\prod_s \mu_s^{n_s} \gamma$, so that the period of $\omega$ on $\prod_s \mu_s^{n_s} \gamma$ is $\prod_s \mu_s^{n_s}$ times the period on $\gamma$. 
Once we work with a homology whose coefficients are rational functions of the $\mu_s$, we can usually find a basis of cycles $\gamma_a$, and 
use the Riemann bilinear identity, with an intersection matrix which will depend on the $\mu_s$. 

As an example, consider the following integral, which leads to the Virasoro-Shapiro amplitude
\begin{equation}
\int_{\mathbb{C}} dz d\bar z |z|^{2A} |1-z|^{2B}
\end{equation}
This integral converges as long as the real parts of $A$ and $B$ are larger than $-1$, and their sum smaller than $-1$. 
The form $\omega$ is now $z^A(1-z)^B dz$, and has monodromies by $\mu_A = e^{2 \pi i A}$ and $\mu_B = e^{2 \pi i B}$
around $0$ and $1$ respectively. These monodromies combine to a monodromy $\mu_A^{-1} \mu_B^{-1}$ around infinity. 
The three ramification points $0$, $1$ and $\infty$ are really on the same footing, and we could move them  to generic positions:
\begin{align}
&\int_{\mathbb{C}} dz d\bar z |z-z_1|^{2A} |z-z_2|^{2B} |z-z_3|^{-2A -2B -4} \cr &= |z_1-z_2|^{2A+2B+2} |z_2-z_3|^{-2A-2}|z_3-z_1|^{-2B-2} \int_{\mathbb{C}} dz d\bar z |z|^{2A} |1-z|^{2B}
\end{align}

\begin{figure}
\includegraphics[width=7cm]{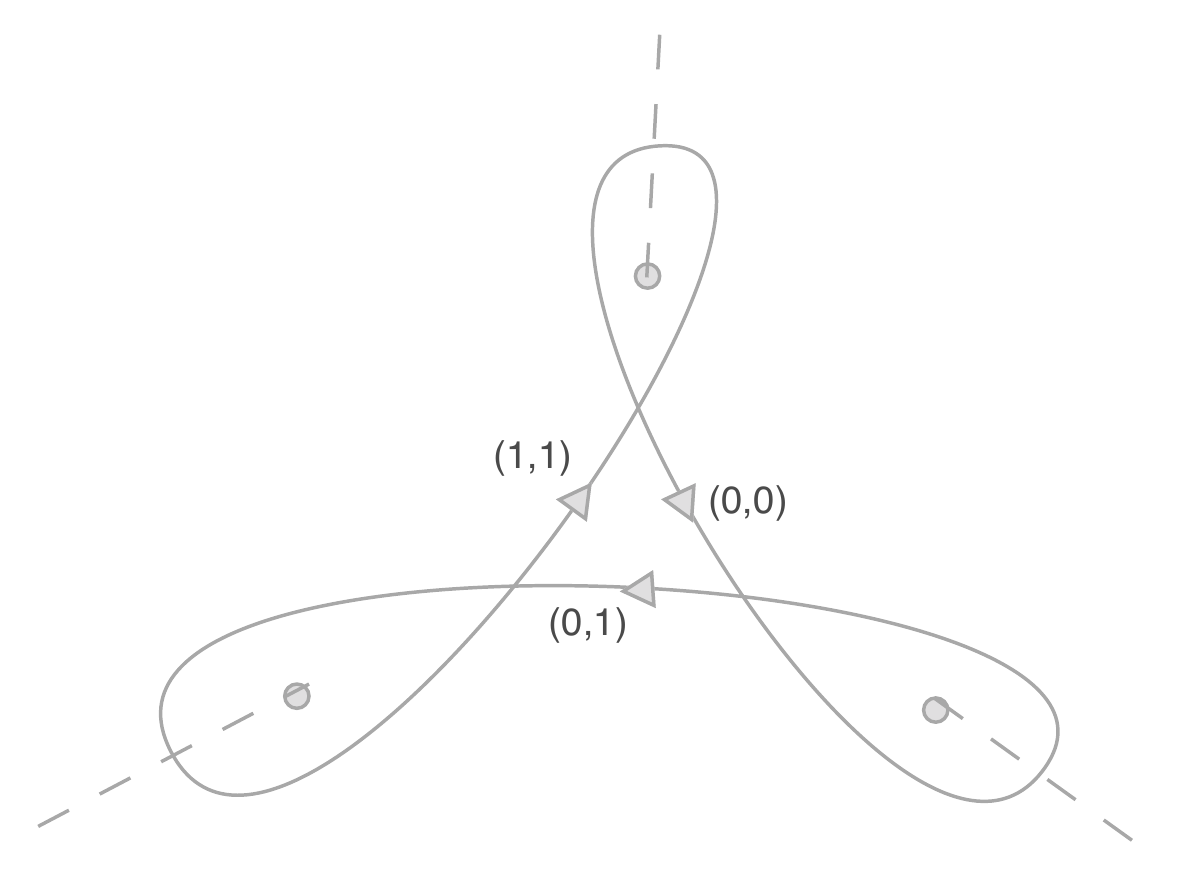}
\includegraphics[width=7cm]{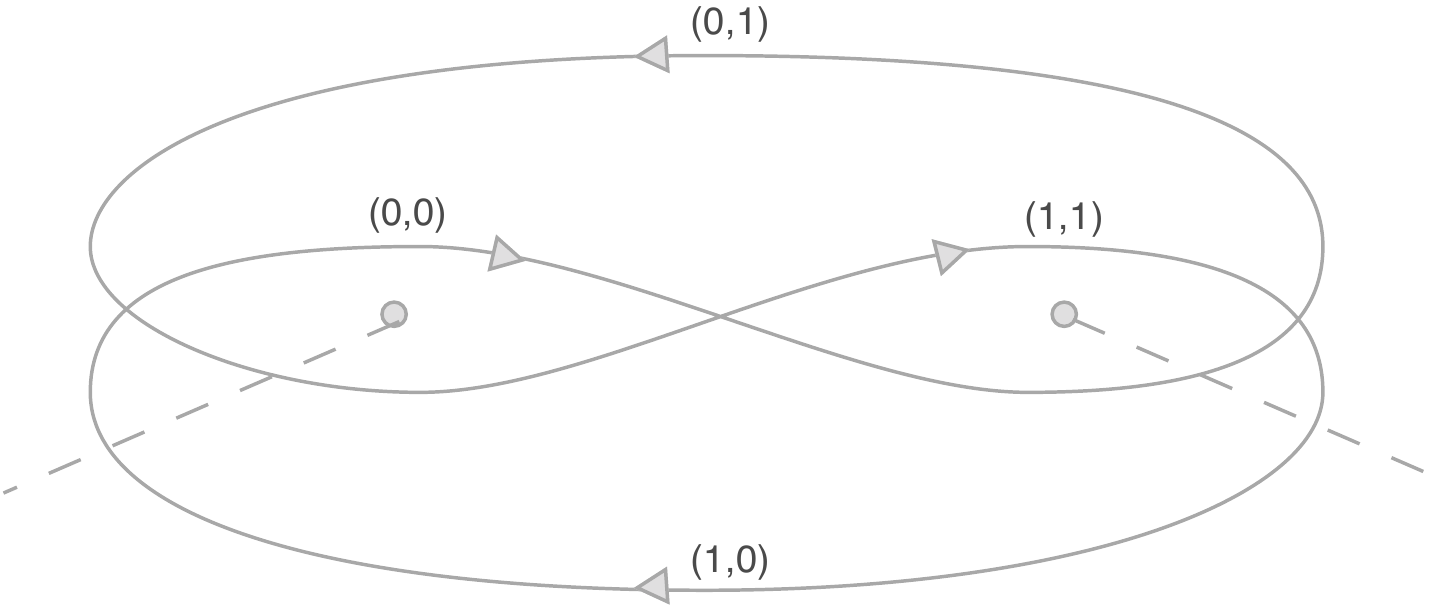}
\caption{\label{fig:poc}Two views of the Pochhammer cycle. On the left: a symmetric presentation. The three punctures are $0$ on the left, $1$ on the right, $\infty$ at the top. We indicate on which sheets the contour runs. On the right, a more traditional presentation, as a cycle wrapping several times around $0$ and $1$. This presentation makes more manifest the relation to an open integration cycle from $0$ to $1$.
A positive intersection point between sheets which differ by $(n_A, n_B)$ units contributes $\mu_A^{n_A} \mu_B^{n_B}-\mu_A^{-n_A} \mu_B^{-n_B}$ to the intersection matrix.}
\end{figure}

There is a single homology generator, which takes the form of a Pochhammer contour $\gamma$, depicted in figure \ref{fig:poc}, together with its $\mu_A^{n_A} \mu_B^{n_B}$ images.
The intersection matrix takes the nice, symmetric form 
\begin{equation}
I = -(1-\mu_A)(1-\mu_B)(1-\mu_A^{-1} \mu_B^{-1})
\end{equation}

The next example, 
\begin{equation}
\int_{\mathbb{C}} dz d\bar z |z|^{2A} |1-z|^{2B} |1-t z|^{2C} 
\end{equation}
or more generically 
\begin{align}
& \int_{\mathbb{C}} dz d\bar z |z-z_1|^{2A} |z-z_2|^{2B}|z-z_3|^{2C} |z-z_4|^{-2A -2B -2C -4} \cr & = |z_1-z_2|^{2A+2B+2} |z_1-z_3|^{2C} |z_4-z_2|^{-2A-2}|z_1-z_4|^{-2B-2C-2} \int_{\mathbb{C}} dz d\bar z |z|^{2A} |1-z|^{2B}|1-t z|^{2C}
\end{align}
where $t = \frac{(z_3 - z_4)(z_1 - z_2)}{(z_3 - z_1) (z_4 - z_2)}$. 

The corresponding one form $\omega$ has monodromies $\mu_A = e^{2 \pi i A}$ around $0$, $\mu_B = e^{2 \pi i B}$ around $1$, $\mu_C = e^{2 \pi i C}$ around $1/t$ and thus 
$\mu_A^{-1}\mu_B^{-1}\mu_C^{-1}$ around infinity. 
The appropriate homology of integration contours has two generators. We can take them to be a Pochhammer cycle around 
$0$ and $1$, and a Pochhammer cycle around $1/t$ and $\infty$. This choice is convenient, as they do not intersect, and we already know their self-intersection.

Furthermore, we can use the well-known integral
\begin{equation}
\int_0^{1} z^A (1-z)^B (1-t z)^C dz = \frac{\Gamma(A+1) \Gamma(B+1)}{\Gamma(A+B+2)} {}_2 F_1(-C,A+1,A+B+2,t)
\end{equation}
or more generally 
\begin{align}
&\int_{z_1}^{z_2} (z-z_1)^A (z-z_2)^B (z-z_3)^C (z-z_4)^{-A-B-C-2} dz = (z_1-z_2)^{A+B+1} (z_1-z_3)^C (z_1-z_4)^{-B-C-1} \cr & (z_4-z_2)^{-A-1}  \frac{\Gamma(A+1) \Gamma(B+1)}{\Gamma(A+B+2)} {}_2 F_1(-C,A+1,A+B+2,\frac{(z_3 - z_4)(z_1 - z_2)}{(z_3 - z_1) (z_4 - z_2)})
\end{align}
which we can specialize to 
\begin{align}
\int_{\infty}^{1/t} z^A (1-z)^{B} (1- t z)^C  dz &=  t^{-A-B-1}(t-1)^{B+C+1}  \frac{\Gamma(C+1) \Gamma(-B-C-A-1)}{\Gamma(-B-A)} \nn
&\qquad\times\quad{}_2 F_1(-A,C+1,-B-C,t)
\end{align}

Thus 
\begin{align}
\int_{\mathbb{C}} dz d\bar z &|z|^{2A} |1-z|^{2B} |1-t z|^{2C} \nn
=& -2 \pi i \frac{\Gamma(A+1) \Gamma(B+1)\Gamma(-A-B-1)}{\Gamma(-A) \Gamma(-B)\Gamma(A+B+2)} |{}_2 F_1(-C,A+1,A+B+2,t)|^2 \cr 
& -2 \pi i \frac{\Gamma(C+1) \Gamma(-A-B-C-1)\Gamma(A+B+1)}{\Gamma(-C) \Gamma(A+B+C+2)\Gamma(-A-B)} |{}_2 F_1(-A,C+1,-B-C,t)|^2 
\end{align}

\subsection{Exponential growth}
The next generalization involves an integral of the form 
\begin{equation}
\int_\Sigma \omega \wedge \tilde \omega e^{W(z) - \bar W(\bar z)}
\end{equation}
where $W$ grows polynomially at infinity. The rapid oscillation makes the integral barely convergent. After the usual analytic continuation, we can improve the behaviour at infinity,
pushing the boundary of the integration cycle towards region where the real part of $W(z)$ grows arbitrarily large and negative, and the real part of $\bar W(\tilde z)$ 
grows arbitrarily large and positive. The integral is then exponentially convergent.

Next, we can try to decompose the integration contour in a convenient basis of appropriate homology of integration cycles for $\omega e^{W(z)}$ and for 
$\tilde \omega e^{- \bar W(\tilde z)}$. The former includes both closed contours and contours which are allowed to end at infinity, in regions where the real part of $W(z)$ grows arbitrarily large and negative. The latter  includes both closed contours and contours which are allowed to end at infinity, in regions where the real part of $\bar W(\tilde z)$ 
grows arbitrarily large and positive.

It is important to observe that there is no well-defined notion of mutual intersection for contours which are allowed to end at infinity, in regions where $\mathrm{Re} W(z) \ll 0$.
Luckily, we do not need that. Rather, there is a well-defined intersection pairing between contours which are allowed to end at infinity, in regions where $\mathrm{Re} W(z) \ll 0$,
and contours which are allowed to end at infinity, in regions where $\mathrm{Re} W(z) \gg 0$
If we pick a set of contours in $\alpha_a$ in the first class, and $\bar \beta_b$ in the second, with intersection $I_{ij}$, 
we can write as usual 
\begin{equation}
\int_\Sigma \omega \wedge \tilde \omega e^{W(z) - \bar W(\bar z)}= - \oint_{\alpha_a} \omega e^{W(z)} I^{-1}_{ab} \oint_{\beta_b} \tilde \omega e^{ - \bar W(\bar z)}
\end{equation}

The first example is a version of the Airy integral
\begin{equation}
\int_{\mathbb{C}} dz d\bar z e^{\frac{z^3}{3} - t z - \frac{ \bar z^3}{3} + \bar t \bar z}
\end{equation}

There are three regions where the contours of integration can end, which are centred around 
rays of phase $\pi/3$, $\pi$, $-\pi/3$. We can denote the integration contours which join consecutive regions counterclockwise 
as $C_1$, $C_2$, $C_3$, with $\sum_i C_i=0$. Dual contours $D_1$, $D_2$, $D_3$ join regions centred around 
rays of phase $-2\pi/3$, $0$, $2\pi/3$. The intersections numbers are basically $I_{i, i+1} = - I_{i,i-1} =1$. 
Thus the integral can be written in the rough form 
$Ai(t) \bar Ai'(\bar t) - Ai'(t) \bar Ai(\bar t)$ where $Ai$ and $Ai'$ are two of the contour integrals.

Another useful example is 
\begin{equation}
\int dz d\bar z |z|^{2A} e^{z - \bar z} 
\end{equation}
which is well-defined before analytic continuation if the real part of $A$ is $-1$. We can improve the behaviour at infinity by 
deforming the contour at large $|z|$ to something like $\tilde z = \bar z + \epsilon |z|$, after which we can allow the real part of $A$ to be bigger than $-1$. 

There is a unique basic integration contour, comes from negative real infinity, goes around the origin counterclockwise, and goes back to negative real infinity. 
A dual contour runs from  positive real infinity and back. The contours have intersection $1-\mu_A$. 
We can write
\begin{equation}
\oint_{\alpha} dz (-z)^A e^{z} = (1-\mu_A) \int_0^{\infty}t^A e^{-t} dt = (1-\mu_A) \Gamma(A+1) 
\end{equation}
and the integral is proportional to $(1-\mu_A)\Gamma(A+1)^2 \sim \frac{\Gamma(A+1)}{\Gamma(-A)}$

In a similar way, one can express the integral 
 \begin{equation}
\int dz d\bar z |z|^{2A} |1 - t z|^{2B} e^{z - \bar z} 
\end{equation}
in terms of confluent hypergeometric functions, and 
\begin{equation}
\int dz d\bar z |1- t z|^{2B} e^{z^2 - \bar z^2} 
\end{equation}
in terms of parabolic cylinder functions.

\subsection{The parabolic cylinder double integral}
\label{app:parabolic cylinder}

Here we calculate the parabolic cylinder double integral 
\begin{align}
\int_{\mb C} dz d\bar z~|z|^{-2(A+1)}e^{-\frac12z^2+\frac12\bar z^2-tz+\bar t \bar z}.
\end{align}
A convenient basis of contours is made of a path from zero to positive infinity and another from negative infinity to zero. A dual basis is a pair of paths from $-i\infty\pm\epsilon$ to $+i\infty\pm\epsilon$. The branch cut is on the negative imaginary axis, such that it splits the dual contours. All the contours can be reduced to the basic integral 
\begin{align}
\int_{0}^\infty dz ~z^{-A-1}e^{-\frac12z^2-tz}=e^{\frac14t^2}\Gamma(-A)D_A(t),
\end{align}
which leads to the result
\begin{align}
\int_{\mb C} dz d\bar z~|z|^{-2(A+1)}e^{-\frac12z^2+\frac12\bar z^2-tz+\bar t \bar z}
=e^{\frac14(t-\bar t)}\Gamma(-A)^2e^{i\frac\pi2A}\l[(D_A(t)\l(D_A(\bar t)-e^{-i\pi A}D_A(-i \bar t)\r)\r.\nn
\qquad\qquad -\l.e^{-i\pi A}D_A(-t)\l(D_A(i\bar t)-e^{i\pi A}D_A(-i\bar t)\r)\r]\nn
=-2i\pi e^{\frac14(t-\bar t)}\gamma(-A)e^{i\frac\pi2A}\l(
D_A(t)D_A(-i\bar t)+i~\gamma(A+1)D_{-A-1}(it)D_{-A-1}(\bar t)\r).
\end{align}
This agrees with the results of section \ref{sec:one-point}.

\section{Direct collision limit of the regular three-point function}
\label{app:one-point}

We rederive the result of section \ref{sec:one-point}, namely the one point-function of a rank two vector, from the direct collision of the regular three-point function. As previously, we assume that the $j^i$ are in the slice $j\in-\tfrac12+i\mb R$. The three-point function of primary fields reads \cite{Ribault:2005wp}
\begin{align}\label{three-point}
\la\Phi_{j^1}(\mu^1|z_1)\Phi_{j^2}(\mu^2|z_2)\Phi_{j^3}(\mu^3|z_3)\ra=&|z_3-z_2|^{-2\Delta_{23}^1}|z_3-z_1|^{-2\Delta_{13}^2}|z_2-z_1|^{-2\Delta_{12}^3}\nn
&\times \delta^{(2)}(\mu^1+\mu^2+\mu^3)D^H\l[\small\begin{array}{lll}j^1&j^2&j^3\\\mu^1&\mu^2&\mu^3\end{array}\r]C^H(j^1,j^2,j^3),
\end{align}
where 
\begin{align}
C^H(j^1,j^2,j^3)=&-\frac{1}{2\pi^3b}\l[\frac{\gamma(b^2)b^{2-2b^2}}{\pi}\r]^{-2-j_{123}}\frac{\Upsilon_b'(0)}{\Upsilon_b(-b(j_{123}+1))}\nn
&\times\frac{\Upsilon_b(-b(2j^1+1))\Upsilon_b(-b(2j^2+1))\Upsilon_b(-b(2j^3+1))}{\Upsilon_b(-bj_{12}^3)\Upsilon_b(-bj_{23}^1)\Upsilon_b(-bj_{31}^2)},\nn
D^H\l[\small\begin{array}{lll}j^1&j^2&j^3\\\mu^1&\mu^2&\mu^3\end{array}\r]=&\pi\frac{|\mu^1|^{4j^1+2}|\mu^2|^{-2j_{13}^2-2}|\mu^3|^{4j^3+2}}{\gamma(-j_{123}-1)}\int_{\mb C} dx d\bar x |x|^{2j_{23}^1}|x+1|^{2j_{12}^3}\l|x-\frac{\mu^1}{\mu^2}\r|^{-2j_{123}-4}.
\end{align}
We use the notation $j_{123}=j^1+j^2+j^3$ and $j_{12}^3=j^1+j^2-j^3$, and the function $\gamma(x)=\Gamma(x)/\Gamma(1-x)$.
The function $\Upsilon_b$ is defined by the integral ($Q=b+b^{-1}$, domain $0<$ Re$(x)<Q$)
\begin{align}
\log \Upsilon_b(x)=\int_0^{\infty}\frac{dt}t\l[\l(\frac Q2-x\r)^2e^{-t}-\frac{\sinh^2[(\frac Q2-x)\frac t2]}{\sinh\frac{bt}2\sinh \frac{t}{2b}}\r].
\end{align}

For the collision of $C^H$ we use the asymptotics ($\tilde \Delta_x=x(Q-x)$)
\begin{align}
\Upsilon_b(x)&\sim \tilde\Delta_x^{-\frac12\tilde\Delta_x+\frac1{12}(1+Q^2)}e^{\frac32\tilde\Delta_x}, 
\end{align}
valid when $x$ has a large imaginary part, with $0<$ Re$(x)<Q$. This leads to the limit
\begin{align}
C^H(j^1,j^2,j^3)\sim&-
2^{(j_0+2)(b^2(j_0+3)+1)}\pi^{j_0-1}b^{(j_0+2)(b^2(j_0+5)-1)}\frac{\gamma(b^2)^{-(j_0+2)}\Upsilon_b'(0)}{\Upsilon_b(-b(j_0+1))}\nn
&\times j_2^{(j_1+3)(b^2(j_0+2)+1)}\l(q_{12}^{-j_{12}^3-1}q_{13}^{-j_{13}^2-1}q_{23}^{-j_{23}^1-1}e^{\frac{j_1^2}{j_2}}
\r)^{2(b^2(j_0+2)+1)}\nn
\sim&-
2^{(j_0+2)(b^2(j_0+3)+1)}\pi^{j_0-1}b^{(j_0+2)(b^2(j_0+5)-1)}\frac{\gamma(b^2)^{-(j_0+2)}\Upsilon_b'(0)}{\Upsilon_b(-b(j_0+1))}\nn
&\times\l(|q_{12}|^{-2(j_{12}^3+1)}|q_{13}|^{-2(j_{13}^2+1)}|q_{23}|^{-2(j_{23}^1+1)}|j_2|^{
(j_0+3)}e^{\frac{j_1^2}{j_2}+\frac{\bar j_1^2}{\bar j_2}}
\r)^{(b^2(j_0+2)+1)},
\end{align}
where we used the equality $j=\bar j$ to relate $q_{ij}$ to $|q_{ij}|$. In $D^H$, the limit of the integral can be evaluated using the change of variable
\begin{align}
x=\frac{\mu^1}{\mu^2}\l(1+i\frac{q_{12}\eta}{\sqrt{2j_2}}\r),
\end{align}
and its anti-holomorphic counterpart. This leads to 
\begin{align}
D^H\l[\small\begin{array}{lll}j^1&j^2&j^3\\\mu^1&\mu^2&\mu^3\end{array}\r]\sim&
|q_{12}q_{13}q_{23}|^{-2j_0-2} 2^{j_0+1}\pi |\mu_2|^{2j_0+2}|j_2|^{j_0+1}\gamma(j_0+2)\nn
&\times\int_{\mc C} d\eta d\bar \eta~|\eta|^{-2(j_0+2)}e^{-\frac12\eta^2+\frac12\bar\eta^2-u\eta+u^*\bar\eta},
\end{align}
with $u$ is the same parameter as defined in section \ref{sec:one-point}. Inserting these results back in eq. (\ref{three-point}), the powers of $q_{ij}$ cancel and we find the limit
\begin{align}
\la \tilde\Phi^{(2)}_{\boldsymbol j}({\boldsymbol \mu}|z)\ra=&
-2^{(j_0+2)(b^2(j_0+3)-1)}\pi^{j_0}b^{(j_0+2)(b^2(j_0+3)-1)} |\mu_2|^{2(j_0+1)}|j_2|^{(j_0+2)(b^2(j_0+3)+2)}\nn
&\times\delta^{(2)}(\mu_0)\gamma(j_0+2)\frac{\gamma(b^2)^{-(j_0+2)}\Upsilon_b'(0)}{\Upsilon_b(-b(j_0+1))}
e^{2i(b^2(j_0+2)+1)\text{Im}\l(\frac{j_1^2}{j_2}\r)}\nn
&\times\int_{\mc C} d\eta d\bar \eta~|\eta|^{-2(j_0+2)}e^{-\frac12\eta^2+\frac12\bar\eta^2-u\eta+u^*\bar\eta}
\end{align}
Using the function $F^{(2)}=j_2^{-(b^2+1)(4j_0+6)}e^{-(b^2+1)j_1^2/j_2}$ to find the one-point function of $\Phi^{(2)}$, we recover the result (\ref{one-pt integral}) found directly for the irregular vector, and fix the overall factor $C(j_0)$.

%\begin{align}
%\la \Phi^{(2)}_{\bf B}({\bf A}|z)\ra=&-\delta^{(2)}(A_1)\l|
%2^{(B_1+2)(b^2(B_1+3)+1)}\pi^{B_1-1}b^{(B_1+2)(b^2(B_1+5)-1)}\frac{\gamma(b^2)^{-(B_1+2)}\Upsilon_b'(0)}{\Upsilon_b(-b(B_1+1))}\r|\nn
%&\times\l|A_3^{B_1+1}B_3^{\frac12(B_1+1)}e^{(b^2(B_1+2)+1)\frac{B_2^2}{B_3}}\r|^2
%\iint d\eta d\bar \eta~\l|\eta^{-(B_1+2)}e^{-\frac12\eta^2+\frac12\bar\eta^2-u\eta+u^*\bar\eta}\r|^2.
%\end{align}

\end{appendix}

\bibliographystyle{JHEP}
\bibliography{irr}

\providecommand{\href}[2]{#2}\begingroup\raggedright\begin{thebibliography}{10}

\bibitem{Ribault:2005wp}
S.~Ribault and J.~Teschner, {\it {H+(3)-WZNW correlators from Liouville
  theory}},  {\em JHEP} {\bf 0506} (2005) 014,
  [\href{http://xxx.lanl.gov/abs/hep-th/0502048}{{\tt hep-th/0502048}}].

\bibitem{Witten:1997sc}
E.~Witten, {\it {Solutions of four-dimensional field theories via M theory}},
  {\em Nucl.Phys.} {\bf B500} (1997) 3--42,
  [\href{http://xxx.lanl.gov/abs/hep-th/9703166}{{\tt hep-th/9703166}}].

\bibitem{Gaiotto:2009hg}
D.~Gaiotto, G.~W. Moore, and A.~Neitzke, {\it {Wall-crossing, Hitchin Systems,
  and the WKB Approximation}},  \href{http://xxx.lanl.gov/abs/0907.3987}{{\tt
  arXiv:0907.3987}}.

\bibitem{Gaiotto:2009we}
D.~Gaiotto, {\it {N=2 dualities}},  {\em JHEP} {\bf 1208} (2012) 034,
  [\href{http://xxx.lanl.gov/abs/0904.2715}{{\tt arXiv:0904.2715}}].

\bibitem{Alday:2009aq}
L.~F. Alday, D.~Gaiotto, and Y.~Tachikawa, {\it {Liouville Correlation
  Functions from Four-dimensional Gauge Theories}},  {\em Lett.Math.Phys.} {\bf
  91} (2010) 167--197, [\href{http://xxx.lanl.gov/abs/0906.3219}{{\tt
  arXiv:0906.3219}}].

\bibitem{Wyllard:2009hg}
N.~Wyllard, {\it {A(N-1) conformal Toda field theory correlation functions from
  conformal N = 2 SU(N) quiver gauge theories}},  {\em JHEP} {\bf 0911} (2009)
  002, [\href{http://xxx.lanl.gov/abs/0907.2189}{{\tt arXiv:0907.2189}}].

\bibitem{Nishioka:2011jk}
T.~Nishioka and Y.~Tachikawa, {\it {Central charges of para-Liouville and Toda
  theories from M-5-branes}},  {\em Phys.Rev.} {\bf D84} (2011) 046009,
  [\href{http://xxx.lanl.gov/abs/1106.1172}{{\tt arXiv:1106.1172}}].

\bibitem{Bonelli:2011jx}
G.~Bonelli, K.~Maruyoshi, and A.~Tanzini, {\it {Instantons on ALE spaces and
  Super Liouville Conformal Field Theories}},  {\em JHEP} {\bf 1108} (2011)
  056, [\href{http://xxx.lanl.gov/abs/1106.2505}{{\tt arXiv:1106.2505}}].

\bibitem{Alday:2010vg}
L.~F. Alday and Y.~Tachikawa, {\it {Affine SL(2) conformal blocks from 4d gauge
  theories}},  {\em Lett.Math.Phys.} {\bf 94} (2010) 87--114,
  [\href{http://xxx.lanl.gov/abs/1005.4469}{{\tt arXiv:1005.4469}}].

\bibitem{Kozcaz:2010yp}
C.~Kozcaz, S.~Pasquetti, F.~Passerini, and N.~Wyllard, {\it {Affine sl(N)
  conformal blocks from N=2 SU(N) gauge theories}},  {\em JHEP} {\bf 1101}
  (2011) 045, [\href{http://xxx.lanl.gov/abs/1008.1412}{{\tt
  arXiv:1008.1412}}].

\bibitem{Pestun:2007rz}
V.~Pestun, {\it {Localization of gauge theory on a four-sphere and
  supersymmetric Wilson loops}},  {\em Commun.Math.Phys.} {\bf 313} (2012)
  71--129, [\href{http://xxx.lanl.gov/abs/0712.2824}{{\tt arXiv:0712.2824}}].

\bibitem{Hama:2012bg}
N.~Hama and K.~Hosomichi, {\it {Seiberg-Witten Theories on Ellipsoids}},  {\em
  Erratum-ibid.} {\bf 1210} (2012) 051,
  [\href{http://xxx.lanl.gov/abs/1206.6359}{{\tt arXiv:1206.6359}}].

\bibitem{Gaiotto:2009ma}
D.~Gaiotto, {\it {Asymptotically free N=2 theories and irregular conformal
  blocks}},  \href{http://xxx.lanl.gov/abs/0908.0307}{{\tt arXiv:0908.0307}}.

\bibitem{Gaiotto:2012sf}
D.~Gaiotto and J.~Teschner, {\it {Irregular singularities in Liouville theory
  and Argyres-Douglas type gauge theories, I}},
  \href{http://xxx.lanl.gov/abs/1203.1052}{{\tt arXiv:1203.1052}}.

\bibitem{Nag1}
H.~{Nagoya} and J.~{Sun}, {\it {Confluent primary fields in the conformal field
  theory}},  {\em Journal of Physics A Mathematical General} {\bf 43} (Nov.,
  2010) 5203, [\href{http://xxx.lanl.gov/abs/1002.2598}{{\tt
  arXiv:1002.2598}}].

\bibitem{Nag2}
H.~{Nagoya} and J.~{Sun}, {\it {Confluent KZ equations for
  $\{$$\backslash$mathfrak $\{$sl$\}$$\}$\_N with Poincar{\'e} rank 2 at
  infinity}},  {\em Journal of Physics A Mathematical General} {\bf 44} (July,
  2011) B5205, [\href{http://xxx.lanl.gov/abs/1002.2273}{{\tt
  arXiv:1002.2273}}].

\bibitem{Nag3}
M.~{Jimbo}, H.~{Nagoya}, and J.~{Sun}, {\it {Remarks on the confluent KZ
  equation for $\backslash$mathfrak$\{$sl$\}$\_2 and quantum Painlev{\'e}
  equations}},  {\em Journal of Physics A Mathematical General} {\bf 41} (May,
  2008) 175205.

\bibitem{FFT}
B.~{Feigin}, E.~{Frenkel}, and V.~{Toledano-Laredo}, {\it {Gaudin models with
  irregular singularities}},  {\em ArXiv Mathematics e-prints} (Dec., 2006)
  [\href{http://xxx.lanl.gov/abs/math/0612}{{\tt math/0612}}].

\bibitem{Alday:2009yn}
L.~F. Alday and J.~Maldacena, {\it {Null polygonal Wilson loops and minimal
  surfaces in Anti-de-Sitter space}},  {\em JHEP} {\bf 0911} (2009) 082,
  [\href{http://xxx.lanl.gov/abs/0904.0663}{{\tt arXiv:0904.0663}}].

\bibitem{Alday:2009ga}
L.~F. Alday and J.~Maldacena, {\it {Minimal surfaces in AdS and the eight-gluon
  scattering amplitude at strong coupling}},
  \href{http://xxx.lanl.gov/abs/0903.4707}{{\tt arXiv:0903.4707}}.

\bibitem{Alday:2009dv}
L.~F. Alday, D.~Gaiotto, and J.~Maldacena, {\it {Thermodynamic Bubble Ansatz}},
   {\em JHEP} {\bf 1109} (2011) 032,
  [\href{http://xxx.lanl.gov/abs/0911.4708}{{\tt arXiv:0911.4708}}].

\bibitem{Alday:2010vh}
L.~F. Alday, J.~Maldacena, A.~Sever, and P.~Vieira, {\it {Y-system for
  Scattering Amplitudes}},  {\em J.Phys.} {\bf A43} (2010) 485401,
  [\href{http://xxx.lanl.gov/abs/1002.2459}{{\tt arXiv:1002.2459}}].

\bibitem{Kanno:2013vi}
H.~Kanno, K.~Maruyoshi, S.~Shiba, and M.~Taki, {\it {W3 irregular states and
  isolated N=2 superconformal field theories}},
  \href{http://xxx.lanl.gov/abs/1301.0721}{{\tt arXiv:1301.0721}}.

\bibitem{CFT}
D.~S. P.~Di~Francesco, P.~Mathieu, {\it {Conformal Field Theory}}, .

\bibitem{Teschner:1997fv}
J.~Teschner, {\it {The Minisuperspace limit of the sl(2,C) / SU(2) WZNW
  model}},  {\em Nucl.Phys.} {\bf B546} (1999) 369--389,
  [\href{http://xxx.lanl.gov/abs/hep-th/9712258}{{\tt hep-th/9712258}}].

\bibitem{Teschner:1997ft}
J.~Teschner, {\it {On structure constants and fusion rules in the SL(2,C) /
  SU(2) WZNW model}},  {\em Nucl.Phys.} {\bf B546} (1999) 390--422,
  [\href{http://xxx.lanl.gov/abs/hep-th/9712256}{{\tt hep-th/9712256}}].

\bibitem{Kanno:2009ga}
S.~Kanno, Y.~Matsuo, S.~Shiba, and Y.~Tachikawa, {\it {N=2 gauge theories and
  degenerate fields of Toda theory}},  {\em Phys.Rev.} {\bf D81} (2010) 046004,
  [\href{http://xxx.lanl.gov/abs/0911.4787}{{\tt arXiv:0911.4787}}].

\bibitem{Dorn:1992at}
H.~Dorn and H.~Otto, {\it {On correlation functions for noncritical strings
  with $c<=1$ $d>=1$}},  {\em Phys.Lett.} {\bf B291} (1992) 39--43,
  [\href{http://xxx.lanl.gov/abs/hep-th/9206053}{{\tt hep-th/9206053}}].

\bibitem{Dorn:1994xn}
H.~Dorn and H.~Otto, {\it {Two and three point functions in Liouville theory}},
   {\em Nucl.Phys.} {\bf B429} (1994) 375--388,
  [\href{http://xxx.lanl.gov/abs/hep-th/9403141}{{\tt hep-th/9403141}}].

\bibitem{Zamolodchikov:1995aa}
A.~B. Zamolodchikov and A.~B. Zamolodchikov, {\it {Structure constants and
  conformal bootstrap in Liouville field theory}},  {\em Nucl.Phys.} {\bf B477}
  (1996) 577--605, [\href{http://xxx.lanl.gov/abs/hep-th/9506136}{{\tt
  hep-th/9506136}}].

\bibitem{Witten:2007td}
E.~Witten, {\it {Gauge theory and wild ramification}},
  \href{http://xxx.lanl.gov/abs/0710.0631}{{\tt arXiv:0710.0631}}.

\bibitem{Harlow:2011ny}
D.~Harlow, J.~Maltz, and E.~Witten, {\it {Analytic Continuation of Liouville
  Theory}},  {\em JHEP} {\bf 1112} (2011) 071,
  [\href{http://xxx.lanl.gov/abs/1108.4417}{{\tt arXiv:1108.4417}}].

\bibitem{Teschner:2001rv}
J.~Teschner, {\it {Liouville theory revisited}},  {\em Class.Quant.Grav.} {\bf
  18} (2001) R153--R222, [\href{http://xxx.lanl.gov/abs/hep-th/0104158}{{\tt
  hep-th/0104158}}].

\bibitem{deBoer:1998pp}
J.~de~Boer, H.~Ooguri, H.~Robins, and J.~Tannenhauser, {\it {String theory on
  AdS(3)}},  {\em JHEP} {\bf 9812} (1998) 026,
  [\href{http://xxx.lanl.gov/abs/hep-th/9812046}{{\tt hep-th/9812046}}].

\bibitem{Maldacena:2000hw}
J.~M. Maldacena and H.~Ooguri, {\it {Strings in AdS(3) and SL(2,R) WZW model
  1.: The Spectrum}},  {\em J.Math.Phys.} {\bf 42} (2001) 2929--2960,
  [\href{http://xxx.lanl.gov/abs/hep-th/0001053}{{\tt hep-th/0001053}}].

\bibitem{Maldacena:2001km}
J.~M. Maldacena and H.~Ooguri, {\it {Strings in AdS(3) and the SL(2,R) WZW
  model. Part 3. Correlation functions}},  {\em Phys.Rev.} {\bf D65} (2002)
  106006, [\href{http://xxx.lanl.gov/abs/hep-th/0111180}{{\tt
  hep-th/0111180}}].

\end{thebibliography}\endgroup

\end{document}